\documentclass[12pt]{article}

\usepackage{amsmath,amsfonts,amssymb,amsthm,latexsym}
\usepackage[mathscr]{eucal}
\usepackage{graphicx} 
\usepackage{textcomp}
\newtheorem{definition}{Definition}
\newtheorem{prop}{Proposition}

\begin{document}

\centerline{\Large \bf  Methods of Modern Differential Geometry}

\centerline{\Large \bf in Quantum Chemistry: TD Theories  }

\centerline{\Large \bf on Grassmann and Hartree-Fock Manifolds}

\bigbreak \bigbreak

\centerline {A. I. Panin}

\centerline{ \sl Chemistry Department, St.-Petersburg State University,}

\centerline {University prospect 26, St.-Petersburg 198504, Russia }

\centerline { e-mail: andrej@AP2707.spb.edu }

\bigbreak

{\bf ABSTRACT: }{\small  Hamiltonian and Schr$\rm \ddot{o}$dinger evolution equations on finite-dimensional projective space are analyzed in detail. Hartree-Fock (HF) manifold is introduced as a submanifold of many electron projective space of states. Evolution equations, exact and linearized, on this manifold are studied. Comparison of matrices of  linearized Schr$\rm \ddot{o}$dinger equations on many electron projective space and on the corresponding HF manifold  reveals the appearance in the HF case a constraining matrix  that includes matrix elements of many-electron Hamiltonian between HF state and double excited determinants. Character of dependence of transition energies on the matrix elements of constraining matrix is established by means of perturbation analysis. It is demonstrated that success of time-dependent HF theory in calculation of transition energies  is mainly due to the wrong behavior of these energies as functions of matrix elements of constraining matrix in comparison with the exact energies.    
}

\bigbreak {\bf Key words: }{\small time-dependent theories; Hamiltonian equations; symplectic geometry}

\bigbreak

\hrule
\bigbreak

{\Large \bf Introduction}

\bigbreak
There exist two widely used by quantum chemists simple approaches for calculation of excitation energies of many electron systems having at their heart the Hartree-Fock (HF) theory.  In both of these approaches it is presupposed that optimal HF molecular spin orbitals (MSOs) are already calculated and that the excited many electron wave functions are linear combinations of  determinants obtained from the HF one by all possible single excitations. The first, most simple, approach is based on the so-called single-excitation configuration interaction (CIS) method and can be technically described as a diagonalization of projection of many electron Hamiltonian on the subspace of single excited determinants. In the second approach, where the so-called time-dependent (TD) method is used, the same subspace of single excited determinants appears as a tangent space to the HF manifold. However, arising on this subspace operator does not coincide with the Hamiltonian projection but involves parameters, accounting  indirectly for the double excitations from the HF state. 

TDHF equations were derived by Dirac from his time-dependent variational principle by constraining trial wave function to remain a single determinant at all times \cite{Dirac}. Interpretation of linearized TDHF equations in terms of harmonic oscillations of a certain fictitious many particle system in a neighborhood of the HF minimum was described by Thouless \cite{Thouless}. More rigorous analysis of TD theories based on methods of modern geometry was undertaken by Rowe and his co-authors \cite{Rowe-1}-\cite{Rowe-4}. He suggested to treat constrained time-dependent many-body quantum mechanics as a Hamiltonian system on a symplectic manifold. This system can be obtained from Dirac's extremal condition of an action integral. 

Our approach, presented here, is closed in its concept to that of Rowe. We, however, simplify  analysis of TD theories by fixing from the very beginning a concrete altlas covering symplectic manifold $\mathscr M$ and work within one selected chart of this atlas centered at the HF optimal determinant. This strategy allows us to use, after realification, the symplectic structure of the parameter space instead of  usually more complicated symplectic structure of a manifold itself. Then  we write down the first order differential equation on a symplectic  parameter space. The right-hand side of this equation is supposed to be a symplectic gradient vector field corresponding to  a local representative of a smooth real-valued  function  on $\mathscr M$. Any such function is called  Hamiltonian function in commonly accepted in theory of symplectic manifolds terminology. The corresponding differential equation is also called Hamiltonian. Subsequent complexification of the realified parameter space leads to the Schr$\rm \ddot{o}$dinger-type evolution equations.

To make our reasoning as independent as possible from numerous textbooks on modern geometry, in Section II necessary definition from manifold theory  together with few simple examples are given. Section III is dedicated to TD theory on finite-dimensional projective spaces. The corresponding linearized Hamiltonian and Schr$\rm \ddot{o}$dinger equations should be considered as  finite-dimensional versions of the exact evolution equations. 
In Section IV HF manifold is introduced and its several atlases are described. Linearized and exact evolution equations on HF manifold for arbitrary quadratic Hamiltonian function are derived. Then the analogous theory is developed for arbitrary (not necessarily quadratic) energy functions depending on matrix elements of 1-density idempotent operators (Hamiltonian functions on Grassmann manifolds). 

Transition energies obtained with the aid of TD theories are usually more close to the exact ones than, say, the CIS energies. In Appendix A it is demonstrated that this is an effect (somewhat paradoxical) of a wrong behavior of TD transitions energies as functions of certain complex parameters in comparison with the behavior of the exact energies. In Appendix B expressions for derivatives of the Gram-Schmidt parametrization function are collected. To the best of our knowledge, this parametrization was introduced in quantum chemistry by Garton \cite{Garton}.                            

\bigbreak \bigbreak

{\Large \bf Basic Definitions} \bigbreak \bigbreak

\bigbreak \bigbreak

Many methods of quantum chemistry are based on a simple idea of
parametrization of a certain subset ${\mathscr M}$ of state
vectors from the $p-$electron sector of the Fock space or  
from the corresponding projective space  by elements of some 
parameter space of a relatively small
dimension with subsequent optimization of chosen parameters using
one or other optimality criterion. A set ${\mathscr M}$ may be a surface or, more generally, a manifold or a variety. We will give all definitions and discuss most important properties of some typical set  ${\mathscr M}$ supposing that  it is either locally Euclidean or locally Hermitian space, which means that each its point has a neighborhood homeomorphic to an open subset of the number space $\mathbb K^n$ where $\mathbb K$ is either the field $\mathbb R$ of real numbers or the field $\mathbb C$ of complex numbers.  

   Let us start with general definitions not presupposing at first, that the set ${\mathscr M}$ is embedded in some space of states. All this definitions may be found in many excellent books dedicated to the manifold theory (see, e.g., \cite{Bourbaki-1}-\cite{Arnold}) including books written by mathematicians especially for physicists \cite{Bruhat, Arnold}. 
\begin{definition}
Triple $c=(U,\varphi,\mathbb K^m)$ is called a chart on ${\mathscr M}$ of dimension $m$ if
(1) $U$ is a subset of ${\mathscr M}$;
(2) $\varphi$ is a bijection of $U$ on an open set in $\mathbb K^m$.
\end{definition}
Subset $U$ is called the domain of the chart $c$ and mapping $\varphi$ is a (local)  coordinate system on ${\mathscr M}$. The inverse of $\varphi$ is called a local parametrization of ${\mathscr M}$. If $x\in U$ then  $c$ is said to be a chart on $\mathscr M$ at point $x$. If $\varphi(x)=0$ then $c$ is called a chart centered at $x$.  

Let $c=(U,\varphi,\mathbb K^m)$ and $c'=(U',\varphi',\mathbb K^{m})$ be two charts on ${\mathscr M}$. These charts are called compatible if 
(1) both $\varphi(U\cap U')$  and $\varphi'(U\cap U')$ are open in ${\mathbb K}^{m}$;
(2) the mappings (that are called transition functions)  $\varphi'\circ \varphi^{-1} :\varphi(U\cap U')\to \varphi'(U\cap U')$ and $\varphi\circ {\varphi'}^{-1} :\varphi'(U\cap U')\to \varphi(U\cap U')$ are $\mathbb K$-analytic.  
 \begin{definition}
Family $\cal A$ of charts $\{(U_i,\varphi_i,\mathbb K^{m})\}_i$ is called an analytic atlas of ${\mathscr M}$ if
(1) ${\mathscr M}=\bigcup\limits_i U_i$;
(2) Any two charts from $\cal A$ are compatible.
\end{definition}
\begin{definition}   
The set $\mathscr M$ with an analytic atlas on it is called $\mathbb K$-analytic manifold.  
\label{atlas}
\end{definition}
Definition \ref{atlas} characterizes manifold with the aid of some its concrete atlas. More elegant definition may be given on the base on the notion of compatible atlases. 
     
Two atlases $\cal A$ and $\cal A'$ of ${\mathscr M}$ are called compatible if ${\cal A}\cup {\cal A'}$ is an atlas of ${\mathscr M}$. It is easy to show that compatibility of atlases is the equivalence relation on the set of all atlases of ${\mathscr M}$ (see, e.g.\cite{Serre}).
\begin{definition}
Class of equivalent atlases defines $\mathbb K$-analytic manifold structure on  ${\mathscr M}$.
\end{definition}

If the ground field is $\mathbb R$ then the requirement of analyticity of transition functions is too restrictive. Instead the notion of differentiable manifolds of class $C^k (k=0,1,\ldots,\infty)$ may be introduced. Transition functions of such manifolds are supposed to be continuously differentiable up to order k. It is clear that transition functions of class $C^k$ are actually $C^k$-diffeomorhisms. Remind as well that $\mathbb R$-analytic manifolds are $C^{\infty}$ ones but not {\it vice versa}.
        
On the same set different manifold structures may
exist. The simplest standard example is the set $\mathbb R$ of real
numbers: two charts $(\mathbb R, id,\mathbb R)$ and $(\mathbb R, \varphi,\mathbb R)$ where $id:x\mapsto x$ and $\varphi:x\mapsto x^3$ determine two different manifold structures on  $\mathbb R$. Indeed, each chart endows $\mathbb R$ with structure of $C^{\infty}$-manifold. But these two charts are not compatible: the transition function $\varphi\circ id^{-1}:x\mapsto x^3$ is smooth and bijective but the inverse mapping $id\circ \varphi^{-1}:x\mapsto x^{\frac{1}{3}}$ is not differentiable at the origin.

The notion of manifold is a very abstract generalization of the classic notions of opened smooth curves and surfaces in Euclidean spaces. If $\sigma:(a,b)\to \mathbb R^m$ is a smooth curve, then it can be interpreted as a smooth 1-dimensional manifold with the chart $c=(\sigma(a,b),\sigma^{-1},\mathbb R)$. Another example is a graphic of smooth function $f:{\mathbb R}^m\to \mathbb R$ (the set of pairs $(x,f(x))$). It is $m$-dimensional manifold ${\mathscr M}_f$ in $\mathbb R^{m+1}$ with atlas consisting of a single chart 
$(\mathscr M_f, (x,x_{m+1})\mapsto x, \mathbb R^m)$. The corresponding parametrization  mapping is $x\mapsto (x,f(x))$. 

Classic ${\mathbb R}$-analytic manifold that can not be covered by a single chart is the unit sphere $\mathscr S^{m-1}$. Among its atlases probably the most simple is constituted by the charts $c_{\alpha\varepsilon}=(U_{\alpha\varepsilon},\varphi_{\alpha\varepsilon},\mathbb R^{m-1})$ where $\varepsilon=\pm 1$, 
\begin{equation}
U_{\alpha \varepsilon}=\{x\in \mathscr S^{m-1}:\varepsilon x_{\alpha}>0\}\\
\end{equation}     
and $\varphi_{\alpha\varepsilon}:x\mapsto (x_1,\ldots,{\hat x}_{\alpha},\ldots,x_m)$. Here the hat over variable means that this variable is omitted.
The inverse mapping (parametrization) is 
\begin{equation}{\gamma}_{\alpha \varepsilon}(x^{(\alpha)})=
\begin{cases} 
x_j &\text{if\,} j\ne\alpha \cr 
\varepsilon \sqrt{1-\|x^{(\alpha)}\|^2} & \text{if\,} j=\alpha \cr 
\end{cases} 
\label{sphpar}
\end{equation}
where $\|x^{(\alpha)}\|<1$.
The minimal number of charts covering $\mathscr S^{m-1}$ is equal to 2. These charts may be constructed with the aid of, say, stereographic projection.

Till now all manifolds we considered were actually  subsets of Euclidean spaces. There arises a natural question: Do there exist manifolds that can not be realized as subsets of appropriate number spaces $\mathbb K^m$ ? In the case of differentiable manifolds the answer is negative. In 1936 Whitney proved that any differentiable manifold of dimension $m$ admits embedding into Euclidean space $\mathbb R^{2m+1}$ \cite{Whitney}. For $\mathbb C$-analytic manifolds, however, the situation is completely different. It is easy to show that compact $\mathbb C$-analytic manifold of positive dimension can not be embedded into Hermitian space $\mathbb C^m$. Instead there exists a broad class of compact $\mathbb C$-analytic manifolds that can be embedded into appropriate projective spaces.  Some of such manifolds that occur in quantum chemistry, will be considered in detail in the next sections.   
  
Let $\mathscr M$ and $\mathscr N$ be two manifolds of dimension $m$ and $n$, respectively,  and $f$ be a mapping $f:\mathscr M\to \mathscr N$. If $(U,\varphi,\mathbb K^m)$ is a chart on $\mathscr M$ and $(V,\psi,\mathbb K^n)$ is a chart on $\mathscr N$ such that $f(U)\subset V$ then the mapping 
$\psi\circ f\circ \varphi^{-1}$ is a classic vector function of $m$ variables defined on a certain open subset of $\mathbb K^m$ and having subset of   $\mathbb K^n$ as its range. This mapping {\it represents} $f$ in the charts under consideration. Mapping $\mathscr M\to\mathscr N$ is called  differentiable at point $x\in \mathscr M$ if its representative in selected charts $(U,\varphi,\mathbb K^m)$ and $(V,\psi,\mathbb K^n)$ is differentiable at  point $\varphi (x)$. It is easy to show that this notion of differentiability does not depend on the choice of charts. Differentiable on $\mathscr M$ functions can therefore be defined as functions differentiable at each point of $\mathscr M$. Mapping $f:\mathscr M\to \mathscr N$ is called morphism of manifolds if any its representative belongs to the class $C^k$ for $C^k$ manifolds and is analytic for analytic manifolds.

Local linearization of a manifold in a neighborhood of some its point leads to very important notion of tangent space. Tangent spaces can be introduced in a number of equivalent ways of which we describe probably the simplest one (see, e.g., \cite{Sternberg}).  Let us consider the set of pairs $(x,\sigma)$ where $x\in \mathscr M$, $\sigma:(-\varepsilon,\varepsilon)\to \mathscr M$ is differentiable mapping (curve) such that $\sigma(0)=x$ (curves passing through a point $x\in \mathscr M$). Two pairs $(x_1,\sigma_1), (x_2, \sigma_2)$ are called equivalent if $x_1=x_2$ and within some chart $c=(U,\varphi,\mathbb K^n)$  on $\mathscr M$ at $x$ the derivatives $D_0[\varphi\circ\sigma_1]$ and $D_0[\varphi\circ\sigma_2]$ coincide. Class $[(x, \sigma)]$ of equivalent pairs is called  tangent vector to $\mathscr M$ at point $x$ and the set of all classes  is called the tangent bundle of manifold $\mathscr M$ and  is denoted ${\sf T}\mathscr M$. To put it more  precisely, ${\sf T}\mathscr M$ is a total space of the tangent bundle which is a triple $(\sf T\mathscr M,\pi,\mathscr M)$ where $\mathscr M$ is a base and $\pi:[(x,\sigma)]\to x$ is a projection of this bundle. Fiber $\pi^{-1}(x)$ over point $x\in\mathscr M$ is called tangent space to $\mathscr M$ at point $x$ and denoted ${\sf T}_x\mathscr M$. It contains classes $[(x,\sigma)]$ with fixed $x$.  
Since each derivative $D_0[\varphi\circ\sigma]$ is, by definition, a linear mapping $\mathbb R\to \mathbb K^m$, we can introduce  a mapping $\theta_c:[(x,\sigma)]\to D_0[\varphi\circ\sigma](1)\in \mathbb K^m$ that is obviously injective. Existence of differentiable curve $\sigma(t)=\varphi^{-1}(\varphi(x)+tv)$ for any vector $v\in \mathbb K^m$ shows that $\theta_c$ is actually a bijection. With the aid of $\theta_c^{-1}$ vector structure from $\mathbb K^m$ is transferred to the tangent space ${\sf T}_x\mathscr M$ which becomes therefore $m$-dimensional vector space. Mapping $\theta_c$ depends, of course, on the chart chosen. Indeed, if at point $x$ another chart $c'=(U',\varphi',\mathbb K^m)$ is taken then $\theta_{c'}=D_{\varphi(x)}[\varphi'\circ\varphi^{-1}]\circ \theta_c$ where, by definition, $D_{\varphi(x)}[\varphi'\circ\varphi^{-1}]$ is an isomorphism of the vector space $\mathbb K^m$. This shows that the transferred vector structure on ${\sf T}_x\mathscr M$ does not depend on the choice of concrete chart. In particular, vector space $\mathbb K^m$ is a manifold that can be covered by a single ('natural`) chart $(\mathbb K^m, id, \mathbb K^m)$. As a representative of class $[(x,\sigma)]$ one can always choose a pair $(x,x+tv)$ where $v=D_0[\varphi\circ\sigma](1)$. We have $\theta_c^{-1}:v\to [(x,x+tv)]$. It is a common practice to identify ${\sf T}\mathbb K^m$ with the direct product $\mathbb K^m\times\mathbb K^m$ and consider the tangent space ${\sf T}_x\mathbb K^m$ as the set of pairs $(x,v)$ where $v$ is a vector outgoing from point $x$ (see, e.g., \cite{Spivak}). 
    
Let $f:\mathscr M\to \mathscr N$ be a morphism of manifolds, $x\in \mathscr M$ and $y=f(x)\in \mathscr N$. Let us suppose that  $c=(U,\varphi,\mathbb K^m)$ and $c'=(V,\psi,\mathbb K^n)$ are  charts on $\mathscr M$ at $x$ and on $\mathscr N$ at $y$, respectively, and $f(U)\subset V$. The derivative of the mapping $F=\psi\circ f\circ \varphi^{-1}$ at the point $\varphi(x)$ is a linear mapping $\mathbb K^m\to \mathbb K^n$. Using the aforementioned bijections $\theta_c:{\sf T}_x\mathscr M\to\mathbb K^m$ and $\theta_{c'}:{\sf T}_{f(a)}\mathscr N\to\mathbb K^n$ we can define linear transformation $\theta_{c'}\circ D_{\varphi(x)}F\circ\theta_c^{-1}$ from ${\sf T}_x\mathscr M$ to ${\sf T}_{f(x)}\mathscr N$ which is called the tangent mapping to $f$ at point $x$ and is denoted  as ${\sf T}_x(f)$. This mapping does not depend on the choice of charts $c$ and $c'$.  In particular, any local parametrization $\gamma:\mathbb K^m\to \mathscr M$ of a certain open subsets of $\mathscr M$  induces tangent mapping ${\sf T}_x(\gamma):{\sf T}_x\mathbb K^m\to{\sf T}_{\gamma(x)}\mathscr M$ at each point of the parameter space. 

In the case of morphism $f:\mathscr M\to \mathbb K^n$ instead of the tangent mapping ${\sf T}_x(f)$ the differential of $f$ is used. Differential is a linear transformation $d_x f:{\sf T}_x\mathscr M\to \mathbb K^n$ that is defined in the following way. If $c$ is a natural chart on $\mathbb K^n$ then $d_x f=\theta_{c}\circ {\sf T}_x(f)$.  It is pertinent to note, however, that the delicate difference between notions of tangent mapping to $f$ and differential of $f$ is usually ignored in mathematical literature and tangent mappings are also called differentials. We will also keep to this tradition. 

If $f$ is a morphism $\mathscr M\to \mathbb K$ then differential $d_x f$ is an element of the vector space ${\sf T}_x\mathscr M^*$ dual to the tangent space to $\mathscr M$ at point $x$. This dual is called the cotangent space and its elements are the so-called co-vectors. 

It is well-known from linear algebra that there is no basis-independent (canonical) isomorphism between a vector space and its dual. An additional algebraic structure on vector space is required to perform canonical transformation of vectors to co-vectors and back. Need in such transformation arises in almost all physical and many mathematical theories. Since transformation co-vector$\to$vector will be used in subsequent sections, we found it reasonable to remind here the necessary definitions.              
 
Let $g:E\times E\to \mathbb R$ be a non-degenerate bilinear form 
on a real vector space $E$. It  defines a
canonical isomorphism of this space and its dual $E^*$. Indeed, for any fixed $v\in E$ the partial mapping   
\begin{equation} g(\cdot ,v):u\to g(u,v)\end{equation}
is a linear functional on $E$ and the mapping 
\begin{equation}\Theta_g :v\to g(\cdot ,v)
 \end{equation} 
is the aforementioned isomorphism. If
$\{e_{\mu}\}_{1\le \mu\le n}$ is a basis of $E$ then the inverse mapping $\Theta_g^{-1}:E^*\to E$ may be written as   
\begin{equation}\Theta_g^{-1} :l\to \sum\limits_{\mu=1}^n e_{\mu} \left [\sum\limits_{\nu=1}^n
g^{\mu\nu}l(e_{\nu})\right ] 
\label{v*v}
\end{equation}
where $g^{\mu\nu}=g^{-1}_{\mu\nu}$  is the matrix inverse to the matrix  
\begin{equation}g_{\mu\nu}=g(e_{\mu},e_{\nu})\end{equation}
of bilinear form $g$ relative to the basis $\{e_{\mu}\}$.

When the ground number field is $\mathbb R$, there are two most important particular cases: bilinear form is non-degenerate symmetric or non-degenerate skew-symmetric. In the first case pair 
$(E,g)$ is called Euclidean space, in the second case it is a symplectic space. Symplectic forms are usually denoted by the symbol $\omega$.

When the ground number field is $\mathbb C$, functionals $l:E\to \mathbb C$ such that
\begin{equation}
l(u+v)=l(u)+l(v)  \ \mbox{and }\  l(cu)=\bar {c}l(u),\quad u,v\in E, c\in \mathbb C
\end{equation} 
are usually considered. The are called $\frac{1}{2}$-linear functional and the vector space (over $\mathbb C$) of such functionals is also denoted $E^*$. Instead of non-degenerate bilinear forms the so-called $1\frac{1}{2}$-linear (or Hermitian) forms  are used. We consider Hermitian forms $\frac{1}{2}$-linear with respect to the first argument:   
\begin{equation} g(u,v) =\sum\limits_{\mu,\nu =1}^q \bar {u}_{\mu}g_{\mu
\nu}v_{\nu} \end{equation}
This is consistent with widely used in physics Dirac's notations: $\langle u|v\rangle$ is $1\frac{1}{2}$-linear form $\frac{1}{2}$-linear with respect to $u$ and $\langle u|$ is just $\frac{1}{2}$-linear functional. Mathematicians usually prefer $1\frac{1}{2}$-linear forms that are $\frac{1}{2}$-linear with respect to the second argument.

Let $\mathscr M$ be a differentiable manifold of dimension $m$ embedded in the Euclidean space $\mathbb R^n$, $f$ be a smooth mapping $\mathbb R^n\to \mathbb R$, and $\gamma :\mathbb R^m\to \mathscr M$ be a parametrization of $\mathscr M$, global if $\mathscr M$ is a surface and local if it is a manifold (without loss of generality one can always take $\mathbb R^m$ as a domain of parametrization mapping). Stationary condition for the function $f\circ\gamma$ at a point $x$ is
\begin{equation}
d_x\left [f\circ\gamma\right ]=d_{\gamma(x)}f\circ d_x\gamma=0
\label{stcond}
\end{equation}
Linear isomorphism $d_x \gamma$ maps $\mathbb R^m$ (parameter space) on the tangent space $T_{\gamma(x)}\mathscr M$ which can be considered as a subspace of $\mathbb R^n$. Image $d_x\gamma(e_i)$ of the canonical basis vectors of the parameters space is a basis of the tangent space $T_{\gamma(x)}\mathscr M$. Differential $d_{a}f$ is a co-vector 
\begin{equation}
d_a f=\sum\limits_{j=1}^n \frac{\partial f}{\partial y_j}(a)dy_j
\end{equation}
where $\{d y_j\}$ is a basis of $(\mathbb R^n)^*$ dual to the canonical basis $\{e_j\}$ of $\mathbb R^n$. In more habitual for physicists Dirac's notations
\begin{equation}
d_a f=\sum\limits_{j=1}^n \frac{\partial f}{\partial y_j}(a)\langle e_j|
\end{equation}
For manifolds embedded in Euclidean spaces the differential of the  mapping $\gamma$ may also be conveniently written  in Dirac's notation as 
\begin{equation}
d_x\gamma=\sum\limits_{i=1}^m |\frac{\partial \gamma}{\partial x_i}(x)\rangle\langle e_i|
\end{equation}
where $\{e_i\}$ is the  canonical basis of the parameter space.
  
Standard Euclidean scalar product $g(e_j,e_{j'})=\langle e_j|e_{j'}\rangle =\delta_{jj'}$ on $\mathbb R^n$ may be used to identify this space and its dual. From Eq.(\ref{v*v}) it readily follows that $\Theta_g^{-1}(\langle e_j|)=|e_j\rangle$ and stationary condition (\ref{stcond}) takes the form
\begin{equation}
\langle \Theta_g^{-1}(d_{\gamma(x)} f)|d_x\gamma(e_i)\rangle =0 \,\,(i=1,\ldots,m)
\label{stcond1}
\end{equation}     
Geometrically this means that gradient of $f$ (vector $\Theta_g^{-1}(d_{\gamma(x)} f)$) at a stationary point $\gamma(x)$ should be perpendicular to the tangent space ${\sf T}_x\mathscr M$. In quantum chemistry conditions of the type of Eq.(\ref{stcond1}) are called Brillouin conditions. 

As has already been mentioned, tangent spaces to manifolds embedded in Euclidean spaces may be considered as a subspaces of the enveloping space and, consequently, they inherit its Euclidean structure (scalar product). In particular, the Gram matrix $G_{ii'}(x)=\langle d_x\gamma(e_i)|d_x\gamma(e_{i'})\rangle$ (overlap matrix in the terminology accepted by quantum chemists) is defined on each tangent space, and,  if smooth in $x$, endows $\mathscr M$ with the structure of Riemannian space. Riemannian metric is used to study the internal geometry of surfaces and  manifolds.
    
Let us return to our simple examples. Differential of the mapping $\gamma:x\to (x,f(x))$ that parametrizes the graphic $\mathscr M_f\subset \mathbb R^{m+1}$ of a smooth function $f:\mathbb R^m\to \mathbb R$ is
\begin{equation}
d_x\gamma=\sum\limits_{i=1}^m |e_i+\frac{\partial f}{\partial x_i}(x)e_{m+1}\rangle\langle e_i|
\end{equation}
Tangent space to $\mathscr M_f$ at point $(x,f(x))$ is spanned by the vectors $d_x\gamma(e_i)=e_i+\frac {\partial f}{\partial x_i}(x)e_{m+1}$ and Gram matrix is $G_{ii'}(x,f(x))=\delta_{ii'}+ \frac {\partial f}{\partial x_i}(x)\frac {\partial f}{\partial x_{i'}}(x)$. Function $f$ may be written as $pr_{m+1}\circ \gamma$ where $pr_{m+1}(x,x_{m+1})=x_{m+1}$. Brillouin conditions for this function $\langle e_{m+1}|d_x\gamma(e_i)\rangle=\frac{\partial f}{\partial x_i}(x)=0$ are just the classic stationary conditions for function $f$ at point $x$.

Differential of the parametrization mapping (\ref{sphpar}) is
\begin{equation}
d{\gamma}_{\alpha \varepsilon}(x^{(\alpha)})=\sum\limits_{j\ne
\alpha}|e_j-\varepsilon \frac{x_j}{\sqrt{1-\|x^{(\alpha)}\|^2}}e_{\alpha}\rangle \langle e_j| 
\end{equation}  
If $f$ is a smooth mapping $\mathbb R^m\to \mathbb R$ then the stationary conditions for its restriction on the unit sphere $\mathscr S^{m-1}$ are
\begin{equation}
\frac{df}{dx}(x)=\lambda x, \, x\in \mathscr S^{m-1}
\end{equation}
where $\frac{df}{dx}(x)=\sum\limits_{j=1}^m \frac{\partial f}{\partial x_j}(x)e_j$ is the gradient of $f$. In the case under consideration Brillouin conditions may be written in the form independent on the chart index.

Let $I$ be an open interval of $\mathbb R$. It is a trivial manifold covered by a single chart $(I,id,\mathbb R)$. Morphism $\sigma:I\to \mathscr M$ is called a smooth curve on $\mathscr M$. Its differential $d_t\sigma$ is a linear mapping $\mathbb R\to {\sf T}_{\sigma(t)}\mathscr M$. Such linear mapping is uniquely determined by the vector $d_t\sigma (1)$ which is called a tangent vector to curve $\sigma$ at point $t\in I$.

Vector field on $\mathscr M$ is defined as a mapping that to each point $x\in \mathscr M$ puts into correspondence a vector from the tangent space ${\sf T}_x\mathscr M$. For example, for any smooth function $f$ on $\mathscr M$  the mapping $x\to \Theta_{g(x)}^{-1}(d_x f)$ is  a vector field on $\mathscr M$ where $g(x)$ is smooth in $x$ non-degenerate bilinear form on ${\sf T}_x\mathscr M$. If $\xi$ is a smooth vector field on $\mathscr M$ then the solution of the first order differential equation 
\begin{equation}
\overset{.}\sigma (t)=\xi(\sigma(t))
\end{equation}
is called an integral curve of this vector field.

In conclusion of this section it must be admitted that in applications the abstract manifold theory recedes in the background and the information given here is therefore somewhat excessive. The role of abstract theory reduces to recognition of geometric object as a manifold and to selection of convenient local coordinates on this manifold. After the concrete atlas of the manifold under consideration is chosen and parametrization mappings are constructed, within a given chart instead of frequently complicated Riemannian or symplectic metrics on the tangent spaces  one can successively use, as a rule much more simple, Euclidean or symplectic structures of the parameter space. Examples of such a strategy are given in the next sections of the present work.

\bigbreak \bigbreak

{\Large \bf Projective Spaces} \bigbreak \bigbreak

\bigbreak \bigbreak

Projective spaces supply us with the simplest example of compact $\mathbb K$-analytic manifolds ($\mathbb K=\mathbb R$ or $ \mathbb C$). From physical viewpoint  state of quantum system is a vector of the relevant Hilbert space determined up to an arbitrary phase prefactor being therefore a point of the corresponding projective space.

The set of 1-dimensional subspaces ('lines passing through the origin`) of the vector space $\mathbb K^{n+1}$ is denoted as  
$\mathbb P_n(\mathbb K)$ or as $\mathbb P(\mathbb K^{n+1})$ and is called the standard $n$-dimensional projective space over the ground  field $\mathbb K$.

For any nonzero $z\in \mathbb K^{n+1}$ symbol $[z]$ stands for 1-dimensional subspace generated by vector $z$. Coordinates $z_0,z_1,\ldots ,z_n$ of vector $z$ are called homogeneous coordinates of line $[z]$ (due to the property $[z]=[\lambda z]$ for any $\lambda \in \mathbb K\backslash \{0\}$). 

For each $\alpha =0,1,\ldots,n$ let us define  
\begin{equation}
U_{\alpha}=\{[z]\in \mathbb P_n(\mathbb  K):z_{\alpha}\ne 0\}
\end{equation}
The mapping 
\begin{equation}
\varphi_{\alpha}:[z_0,\ldots,z_{\alpha -1},z_{\alpha},z_{\alpha +1},\ldots, z_n]\mapsto (\frac{z_0}{z_{\alpha}},\ldots,\frac{z_{\alpha -1}}{z_{\alpha}},\frac{z_{\alpha +1}}{z_{\alpha}},\ldots, \frac{z_n}{z_{\alpha}})\in \mathbb K^n
\label{h_a}
\end{equation}
is a local coordinate system on $U_{\alpha}$ and the family of charts $c_{\alpha}=(U_{\alpha},\varphi_{\alpha},\mathbb K^n)$ 
is an atlas of $\mathbb K$-analytic structure on $\mathbb P_n(\mathbb K)$. The inverse mapping
\begin{equation}
\varphi^{-1}_{\alpha}:(\zeta_0,\ldots,{\hat \zeta}_{\alpha},\ldots,\zeta_n)\mapsto [\zeta_0,\ldots,1,\ldots,\zeta_n]
\label{projpar1}
\end{equation}
is a local parametrization of $U_{\alpha}$ by elements of $\mathbb K^n$. The hat over variable means that this variable should be omitted. 

There exists  surjective mapping $\pi:\mathbb K^{n+1}\backslash \{0\} \to \mathbb P_n(\mathbb K)$ defined by the relation $\pi(z)=[z]$. Its restriction to the unit sphere $S^n$ for $\mathbb K=\mathbb R$ and to the unit sphere  
\begin{equation}
S^{2n+1}=\{z\in \mathbb C^{n+1}:\langle z|z\rangle =\sum\limits_{j=0}^n \bar {z}_jz_j=1\}
\end{equation}
for $\mathbb K=\mathbb C$ is also surjective. Since for any nonzero $z\in \mathbb K^{n+1}$ 
\begin{equation}
[z]\cap S^n=\{+\frac{z}{\|z\|},-\frac{z}{\|z\|}\}\quad \text{if}\quad \mathbb K=\mathbb R
\end{equation}
and
\begin{equation} 
[z]\cap S^{2n+1}=\{e^{{\rm i}\varphi}\frac{z}{\|z\|}:\varphi\in [0,2\pi)\}\quad \text{if}\quad \mathbb K=\mathbb C 
\end{equation}
it is possible to realize the projective space $\mathbb P_n(\mathbb K)$ either as a quotient of the unit sphere $S^n$ modulo the equivalence relation 
\begin{equation}
z\sim z'\Leftrightarrow z'=\pm z 
\end{equation}
($\mathbb K=\mathbb R$), or as a quotient of the unit sphere
$S^{2n+1}$ modulo the equivalence relation 
\begin{equation}
z\sim z'\Leftrightarrow z'=e^{{\rm i}\varphi}z
\label{eqrel}
\end{equation}
($\mathbb K=\mathbb C$).   
In further discussion we confine ourselves to the most interesting for us case $\mathbb K=\mathbb C$. It is easy to see that the mapping 
\begin{eqnarray}
\zeta=(\zeta_0,\ldots,{\hat \zeta}_{\alpha},\ldots,\zeta_n)\mapsto (\zeta_0,\ldots,1,\ldots,\zeta_n)\qquad\qquad\qquad\nonumber\\
\mapsto\frac{1}{\sqrt{1+\langle \zeta|\zeta\rangle}}(\zeta_0,\ldots,1,\ldots,\zeta_n)
\mapsto \bigcup_{\varphi}\left \{
\frac{e^{{\rm i}\varphi}}{\sqrt{1+\langle \zeta|\zeta\rangle}}(\zeta_0,\ldots,1,\ldots,\zeta_n)\right \}
\end{eqnarray} 
is a local parametrization of $\mathbb P_n(\mathbb C)$ realized as  a quotient of $S^{2n+1}$ modulo the equivalence relation (\ref{eqrel}). Vector $\frac{1}{\sqrt{1+\langle \zeta|\zeta\rangle}}(\zeta_0,\ldots,1,\ldots,\zeta_n)$ is a representative of the corresponding equivalence class. In fact, we have $\mathbb R$-analytic local parametrization
\begin{equation}
\gamma_{\alpha}:(\zeta_0,\ldots,{\hat \zeta}_{\alpha},\ldots,\zeta_n)\mapsto \frac{1}{\sqrt{1+\langle \zeta|\zeta\rangle}}(\zeta_0,\ldots,1,\ldots,\zeta_n)
\label{reppar}
\end{equation}
of the representatives of the equivalence classes (\ref{eqrel}) and $\gamma_{\alpha}(\mathbb C^n)$ is a $2n$-dimensional surface situated on  $S^{2n+1}$.   

For geometric characterization of tangent spaces to the projective manifold $\mathbb P_n(\mathbb C)$ it seems reasonable to start with the realification of the complex vector space $\mathbb C^{n+1}$ (see \cite{Arnold}) that gives the real vector space $\mathbb C_{\mathbb R}^{n+1}\sim \mathbb R^{2n+2}$ with the standard basis 
\begin{equation}
e_0,e_1,e_2,\ldots,e_n,{\rm i}e_0,{\rm i}e_1,{\rm i}e_2,\ldots,{\rm i}e_n
\label{canbas}
\end{equation}
Hermitian scalar product on $\mathbb C^{n+1}$ (which is supposed to be $\frac{1}{2}$-linear with respect to the first argument)  may be written as 
\begin{equation} 
\langle z|z'\rangle = (z,z')+{\rm i}[z,z']
\end{equation}
where 
\begin{equation}
(z,z')={\rm Re}\ \langle z|z'\rangle
\label{Euclid}
\end{equation}
is the Euclidean scalar product and
\begin{equation}
[z,z']={\rm Im}\ \langle z|z'\rangle
\label{symplectic}
\end{equation}
is the symplectic one, both of them are non-degenerate. Basis (\ref{canbas}) is orthonormal with respect to the Euclidean scalar product (\ref{Euclid}) and is also standard symplectic basis: $[e_i,e_j]=[{\rm i}e_i,{\rm i}e_j]=0$, and $[e_i,{\rm i}e_j]=\delta_{ij}$. Euclidean and symplectic scalar products are connected by the relation $[z,z']=({\rm i}z,z')$.    

Tangent vector space to $S^{2n+1}$ at some point $z$ is the orthogonal complement to vector $z$ in $\mathbb R^{2n+2}$ with respect to the Euclidean scalar product (\ref{Euclid}):
\begin{equation}
{\sf T}_z S^{2n+1}=\{\zeta\in \mathbb C_{\mathbb R}^{n+1}:(z,\zeta)=0 \}
\end{equation}
Point $z\in S^{2n+1}$ is a representative of the circle $\{e^{{\rm i}\varphi}z:\varphi\in[0,2\pi)\}$. Tangent vector to this circle at point $z$ is $\frac{d}{d{\varphi}}e^{{\rm i}\varphi}z|_{\varphi=0}={\rm i }z$. It is easy to ascertain that
\begin{equation}
\langle z|\zeta\rangle =0 \Leftrightarrow (z,\zeta)=0\ \text{and}\ ({\rm i}z,\zeta)=0
\end{equation} 
As a result, the tangent space ${\sf T}_{[z]}\mathbb P_n(\mathbb C)$ at point $[z]$ is isomorphic to $(\mathbb C z)^{\perp}$ (the orthogonal complement to the line $\mathbb C z$ with respect to the Hermitian scalar product on $\mathbb C^{n+1}$). Orthogonal projection of arbitrary vector $\xi\in \mathbb C^{n+1}$ on the orthogonal complement to $z$ is of the form 
\begin{equation}
\xi - \frac{\langle z|\xi \rangle }{\langle z|z\rangle}z
\end{equation}
If $d_z \pi:{\sf T}_z\mathbb C^{n+1} \to {\sf T}_{[z]}\mathbb P_n(\mathbb C)$ is the differential of $\pi$ at $z$, and $\eta_1=d_z \pi (\xi_1)$, $\eta_2=d_z \pi (\xi_2)$ are two tangent vectors from ${\sf T}_{[z]}\mathbb P_n(\mathbb C)$ (their concrete nature is irrelevant) then it is possible to introduce on  ${\sf T}_{[z]}\mathbb P_n(\mathbb C)$ the following Hermitian scalar product 
\begin{equation}
\langle \eta_1|\eta_2 \rangle _{[z]}=\frac{\langle \xi_1|\xi_2\rangle\langle z|z\rangle-\langle \xi_1|z\rangle\langle z|\xi_2\rangle}{\langle z|z\rangle ^2}
\label{hermit1}
\end{equation}

Now let us consider parametrization (\ref{reppar}) assuming that for each $\alpha$ the parameter space $\mathbb C^n$ is embedded in $\mathbb C^{n+1}$: $\mathbb C^n\subset \mathbb C^n\oplus\mathbb Ce_{\alpha}$. Differential  $d_{\zeta}\gamma_{\alpha}$ is an isomorphism 
${\sf T}_{\zeta}\mathbb R^{2n} \to {\sf T}_{\gamma_{\alpha}(\zeta)}\gamma_{\alpha}(\mathbb R^{2n})$. Simple calculations give
\begin{equation}
d_{\zeta}\gamma_{\alpha}(\xi)=\frac{1}{[1+\langle \zeta |\zeta \rangle]^{\frac{1}{2}}}\left [\xi - (\xi, \gamma_{\alpha}(\zeta))\gamma_{\alpha}(\zeta)\right ]
\label{imksi}
\end{equation}  
where $\xi$ belongs to ${\sf T}_{\zeta}\mathbb R^{2n}\sim\mathbb R^{2n}$ and $(\gamma_{\alpha}(\zeta),\xi)$ is the Euclidean scalar product (\ref{Euclid}). It is easy to see that with respect to this scalar product  vectors (\ref{imksi})  are orthogonal to $\gamma_{\alpha}(\zeta)$ and to ${\rm i}e_{\alpha}$. Thus, with such an approach,
\begin{equation}
{\mathbb R}^{2n+2}={\sf T}_{\gamma_{\alpha}(\zeta)}S^{2n+1}\oplus{\mathbb R}\gamma_{\alpha}(\zeta)
\end{equation}
and
\begin{equation}
{\sf T}_{\gamma_{\alpha}(\zeta)}S^{2n+1}={\sf T}_{\gamma_{\alpha}(\zeta)}\gamma_{\alpha}(\mathbb R^{2n})\oplus\mathbb R {\rm i}e_{\alpha} 
\end{equation}
The expressions for the Euclidean and symplectic scalar products on the tangent space ${\sf T}_{\gamma_{\alpha}(\zeta)}\gamma_{\alpha}(\mathbb R^{2n})$ are 
\begin{equation}
g^{(\alpha)}_{\zeta}(\eta_1,\eta_2)= \frac{(\xi_1,\xi_2)(1+\|\zeta\|^2)-(\xi_1,\zeta)(\zeta,\xi_2)}{(1+\|\zeta\|^2)^2}
\label{gform}
\end{equation}
and
\begin{equation}
\omega^{(\alpha)}_{\zeta}(\eta_1,\eta_2)=\frac{[\xi_1,\xi_2](1+\|\zeta\|^2)-(\xi_1,\zeta)[\zeta,\xi_2]-[\xi_1,\zeta](\zeta,\xi_2)}{(1+\|\zeta\|^2)^2}
\label{omegaform}
\end{equation}
where  $\eta_1=d_{\zeta}\gamma_{\alpha}(\xi_1)$ and $\eta_2=d_{\zeta}\gamma_{\alpha}(\xi_2)$.  Basis vectors of the tangent space ${\sf T}_{\gamma_{\alpha}(\zeta)}\gamma_{\alpha}(\mathbb R^{2n})$ are just the images of basis vectors from the parameter space. We see that these scalar products are rather complicated in comparison with the analogous scalar products on the parameter space.

Now let us consider a quadratic 'energy` function on the projective space $\mathbb P_n(\mathbb C)$ 
\begin{equation}
E([z])=\frac{1}{2}\frac{\langle z|H|z\rangle}{\|z\|^2}
\label{efunc1}
\end{equation}
that can be locally presented as 
\begin{equation}
E(\zeta)=\frac{1}{2}\langle \gamma_{\alpha}(\zeta)|H|\gamma_{\alpha}(\zeta)\rangle
\label{efunc2} 
\end{equation}
where $H$ is some Hermitian operator on $\mathbb C^{n+1}$, and suppose that $H$ has $e_{\alpha}$ as its non-degenerate eigenvector. $E(\zeta)$ is a  representation of the function (\ref{efunc1}) within the chart $c_{\alpha}=(U_{\alpha}, \varphi_{\alpha},\mathbb R^{2n})$. Standard Euclidean and symplectic inner products defined by Eqs.(\ref{Euclid})-(\ref{symplectic}) exist  on the parameter space $\mathbb R^{2n}$. Since basis $\{e_j,{\rm i}e_j\},j\ne \alpha$ is symplectic, matrix of the symplectic form (\ref{symplectic}) is 
\begin{equation}
\Omega^{(\alpha)}=\begin{pmatrix}
0&{\rm I}_n\\
-{\rm I}_n&0
\end{pmatrix}
\end{equation}
Without loss of generality in the remainder of this section we assume that $\alpha =0$ and suppress index $\alpha$ in all forthcoming expressions. 

After realification the energy function $E(x,y)$ becomes
\begin{equation}
E(x,y)=\frac{E_0+\sum\limits_{i,j=1}^nx_ix_jA_{ij}+\sum\limits_{i,j=1}^ny_iy_j{\rm A}_{ij}-2\sum\limits_{i,j=1}^nx_iy_j{\rm B}_{ij}}{2(1+\|x\|^2+\|y\|^2)}
\label{efunc3}
\end{equation}     
where $E_0$ is the eigenvalue of $H$ corresponding to the eigenvector $e_0$, and $\rm A=Re\ H$, $\rm B= Im\  H$ are real and imaginary components of the operator $H$ matrix with respect to the basis (\ref{canbas}), $\rm A$ is symmetric and $\rm B$ is skew-symmetric.  
Partial first derivatives of $E(x,y)$ are
\begin{subequations}
\begin{eqnarray}
\frac{\partial}{\partial x_j}E(x,y)=-\frac{2x_jE(x,y)-\sum\limits_{i=1}^nA_{ji}x_i+\sum\limits_{i=1}^nB_{ji}y_i}{1+\|x\|^2+\|y\|^2}\\
\frac{\partial }{\partial y_j}E(x,y)=-\frac{2y_jE(x,y)-\sum\limits_{i=1}^nA_{ji}y_i-\sum\limits_{i=1}^nB_{ji}x_i}{1+\|x\|^2+\|y\|^2}
\end{eqnarray}
\end{subequations}
Realified matrix of the second derivatives  at the origin $(x,y)=(0,0)$ is
\begin{equation}
{\mathscr H}=\begin{pmatrix}
\ {\rm A}-E_0{\rm I}_n&-{\rm B}\\
{\rm B}&\ {\rm A}-E_0{\rm I}_n
\end{pmatrix}
\end{equation}
Spectra of operator $H-E_0I_n$ in the Hermitian space $\mathbb C^{n}$ and the Hessian $d^2_{(0,0)}E$ in the Euclidean  space $\mathbb R^{2n}$ are identical. Indeed, it is easy to show that 
\begin{equation}
\det\left [{\mathscr H}-\omega {\rm I}_{2n}\right ]=|\det\left [{\rm H}-(E_0+\omega){\rm I}_n\right ]|^2
\end{equation}
Eigenvalues are just the 'transition energies` $\omega_{i0}=E_i-E_0$. Positive definiteness of Hessian implies that the function (\ref{efunc1}) has its minimum at the point $e_0$.
 
Now let us try to exploit the symplectic structure of the parameter space. In theory of symplectic manifolds any smooth real-valued function  on a symplectic manifold is called a Hamiltonian function. In particular, both function (\ref{efunc1}) and its local representative (\ref{efunc2}) are the Hamiltonian ones. 

In general case, a  symplectic manifold is a pair $({\mathscr M},\omega)$ where ${\mathscr M}$ is an even-dimensional differentiable manifold and  $\omega$ is closed skew-symmetric 2-form on ${\mathscr M}$ that is, for each $x \in {\mathscr M}$ the mapping $\omega_x:{\sf T}_x{\mathscr M}\times{\sf T}_x{\mathscr M}\to \mathbb R$ is non-degenerate bilinear skew-symmetric,  $\omega_x$ varies smoothly in $x$, and $d\omega=0$ ($d$ is the exterior derivative of $\omega$). For example, the mapping $\zeta\to \omega^{(\alpha)}_{\zeta}$ (see Eq.(\ref{omegaform})) endows the surface $\gamma_{\alpha}(\mathbb R^{2n})$ with the symplectic structure.

For any smooth function $f:{\mathscr M}\to \mathbb R$ its differential at point $x\in {\mathscr M}$ is a covector $d_xf:{\sf T}_x{\mathscr M}\to \mathbb R$. The image of $d_xf$ with respect to the  isomorphism $\Theta_{\omega}^{-1}$ (see Eq.(\ref{v*v})) is a vector of the tangent space ${\sf T}_x{\mathscr M}$ and the first order differential equation
\begin{equation}
\frac{d}{dt}\gamma (t)= \Theta_{\omega}^{-1}\left (d_{\gamma(t)}f\right )  
\label{Heq}
\end{equation}
is called a Hamiltonian one. Critical points of $f$ are just the singular points of the vector field $x\to \Theta_{\omega}^{-1}\left (d_xf\right )$ and {\it vice versa}. 

In certain  situations to study the stability of solution of Eq.(\ref{Heq}) in a neighborhood of some its singular point it is sufficient to analyze the linearization of this differential equation (see, e.g. \cite{Arnold}). 

For the  energy function (\ref{efunc3}) its differential is a covector
\begin{equation}
d_{(x,y)}E=\sum\limits_{j=1}^n \frac{\partial}{\partial x_j}E(x,y)dx_j+\sum\limits_{j=1}^n \frac{\partial}{\partial y_j}E(x,y)dy_j
\end{equation}
that can be transformed to the the symplectic gradient to give 
\begin{equation}
\Theta_{\omega}^{-1}(d_{(x,y)}E)=-\sum\limits_{j=1}^ne_j\frac{\partial}{\partial y_j}E(x,y)+\sum\limits_{j=1}^n
{\rm i}e_j\frac{\partial}{\partial x_j}E(x,y)
\label{v*tovd}
\end{equation}
(here the standard symplectic structure of the parameter space is used).

The Hamiltonian equations in coordinate form are
\begin{subequations}
\begin{eqnarray}
\overset{.}x_j=&-\frac{\partial}{\partial y_j}E(x,y)\\
\overset{.}y_j=&\frac{\partial}{\partial x_j}E(x,y)
\label{Hequ}
\end{eqnarray}
\end{subequations}
where $j=1,2,\ldots,n$.

Linearization of Hamiltonian equations  in a neighborhood of its critical point means that Hamiltonian function is replaced by its quadratic approximation. For $E(x,y)$ in a neighborhood of the origin we have
\begin{eqnarray}
E^{(2)}(x,y)=\nonumber\qquad\qquad\qquad\qquad\qquad\qquad\\
E_0+\frac{1}{2}\left [\sum\limits_{i,j=1}^nx_ix_j({\rm A}-E_0{\rm I}_n)_{ij} + \sum\limits_{i,j=1}^ny_iy_j({\rm A}-E_0{\rm I}_n)_{ij}-2\sum\limits_{i,j=1}^nx_iy_j{\rm B}_{ij}\right ]
\end{eqnarray}   
The corresponding linearized Hamiltonian equations are
\begin{subequations}
\begin{eqnarray}
\overset{.}x_j=&-\sum\limits_{j=1}^n({\rm A}-E_0{\rm I}_n)_{ij}y_j-\sum\limits_{j=1}^n{\rm B}_{ij}x_j\nonumber\\
\overset{.}y_j=&\sum\limits_{j=1}^n({\rm A}-E_0{\rm I}_n)_{ij}x_j-\sum\limits_{j=1}^n{\rm B}_{ij}y_j
\end{eqnarray}
\end{subequations}
or, in a matrix form
\begin{equation}
\left (\begin{array}{c}
\overset{.}x\\
\overset{.}y
\end{array}\right )
=-\Omega{\mathscr H}
\left (\begin{array}{c}
x\\
y
\end{array}\right )
\label{matHequ}
\end{equation}

Characteristic roots of real matrix $-\Omega {\mathscr H}$ are  purely imaginary. Indeed, 
\begin{equation}
\det\left [\Omega{\mathscr H}+\lambda {\rm I}_{2n}\right ]=|\det\left [{\rm H}-(E_0+{\rm i}\lambda ){\rm I}_n\right ]|^2 
\end{equation}
and, consequently, the aforementioned roots are $\pm {\rm i}\omega_{i0}$. 

Realified parameter space $(\mathbb C^n)_{\mathbb R}\sim \mathbb R^{2n}$ can be again complexified to give complex vector space $\mathbb C\otimes_{\mathbb R} \mathbb R^{2n}$ of complex dimension $2n$ with basis vectors $\{1\otimes e_j,1\otimes {\rm i}e_j\}$ where $1$ is the unit of $\mathbb C$. Hermitian structure on this space is introduced in the following  way: for any $\alpha, \alpha' \in \mathbb C$ and any $v,v'\in \mathbb R^{2n}$
\begin{equation}
\langle \alpha\otimes v|\alpha'\otimes v'\rangle={\bar \alpha}\alpha'(v,v')
\label{hermit2}
\end{equation}
where $(\cdot,\cdot)$ is standard Euclidean scalar product on $\mathbb R^{2n}$. Basis $\{1\otimes e_j,1\otimes {\rm i}e_j\}$ is orthonormal with respect to this scalar product. 
  
Let us select in $\mathbb C\otimes_{\mathbb R} \mathbb R^{2n}$ a new orthogonal basis
\begin{equation}
f_j=\frac{1}{2}\left [(1\otimes e_j)-{\rm i}(1\otimes {\rm i}e_j)\right ],\quad \bar{f_j}=\frac{1}{2}\left [(1\otimes e_j)+{\rm i}(1\otimes {\rm i}e_j)\right ] 
\label{fbas}
\end{equation}
The corresponding transformation matrix is 
\begin{equation}
{\rm C}=\begin{pmatrix}
\ \ \ {\rm I}_n&{\rm I}_n\\
-{\rm i}{\rm I}_n&{\rm i}{\rm I}_n
\end{pmatrix}
\end{equation}
In this basis the Hamiltonian equations (\ref{Hequ}) take the Schr$\rm \ddot{o}$dinger-type form
\begin{subequations}
\begin{eqnarray}
\overset{.}\zeta_j=&2{\rm i}\frac{\partial}{\partial {\bar \zeta}_j}E(\zeta,\bar \zeta)\nonumber\\
\overset{.}{\bar \zeta}_j=&-2{\rm i}\frac{\partial}{\partial \zeta_j}E(\zeta,\bar \zeta)
\end{eqnarray}
\label{Schrod1}
\end{subequations}
where ${\zeta}_j=x_j+{\rm i}y_j$ and where, by definition, 
\begin{subequations}
\begin{eqnarray}
\frac{\partial}{\partial \zeta_j}=\frac{1}{2}\left (\frac{\partial}{\partial x_j}-{\rm i}\frac{\partial}{\partial y_j}\right )\\
\frac{\partial}{\partial {\bar\zeta}_j}=\frac{1}{2}\left (\frac{\partial}{\partial x_j}+{\rm i}\frac{\partial}{\partial y_j}\right )
\end{eqnarray}
\label{derz}
\end{subequations}
On the complexified parameter space $\mathbb C\otimes_{\mathbb R} \mathbb R^{2n}$ energy can be considered as a function of {\it independent} variables $\zeta,\bar \zeta$.     

Linearized matrix Hamiltonian equation (\ref{matHequ}) in variables $\zeta,\bar\zeta$ becomes
\begin{equation}
\left (\begin{array}{c}
\overset{.}\zeta\\
\overset{.}{\bar\zeta}
\end{array}\right )
=-{\rm C}^{-1}\Omega{\mathscr H}{\rm C}
\left (\begin{array}{c}
\zeta\\
\bar\zeta
\end{array}\right )
\label{Schrod2}
\end{equation}
where
\begin{equation}
{\rm C}^{-1}\Omega{\mathscr H}{\rm C}=-{\rm i}\begin{pmatrix}
{\rm H}-E_0{\rm I}_n&0\\
0&-\left ({\rm \bar H}-E_0{\rm I}_n\right )
\end{pmatrix}
\end{equation}
Spectra of real Hamiltonian matrix $-\Omega{\mathscr H}$ and  complex Schr$\rm \ddot{o}$dinger-type matrix $-{\rm C}^{-1}\Omega{\mathscr H}{\rm C}$ are  obviously identical. It is clear as well that the last matrix is diagonalizable over $\mathbb C$. This means, in particular, that real non-symmetric matrix $-\Omega {\mathscr H}$ is also diagonalizable over ${\mathbb C}$. Over the field $\mathbb R$ of real numbers this matrix can be transformed into block-diagonal form with $n$ skew-symmetric $2\times 2$ blocks. Indeed, if $u_i$ is an eigenvector  of matrix $-\Omega {\mathscr H}$ belonging to the eigenvalue ${\rm i}\omega_{i0}$ then $\bar {u}_i$ is also the eigenvector of this matrix belonging to the eigenvalue $-{\rm i}\omega_{i0}$. It is easy to show that vectors ${\rm Re}\,u_i$ and ${\rm Im}\,u_i$ constitute a basis of two-dimensional invariant subspace of matrix $-\Omega {\mathscr H}$ in real parameter space $\mathbb R^{2n}$. In this basis matrix $-\Omega {\mathscr H}$ becomes a direct sum of $2\times 2$ real matrices and differential equation (\ref{matHequ}) becomes a direct product of $n$ equations
\begin{equation}
\begin{pmatrix}
\overset{.}p_i\\
\overset{.}q_i
\end{pmatrix}=\begin{pmatrix}
0&\omega_{i0}\\
-\omega_{i0}&0
\end{pmatrix}
\begin{pmatrix}
p_i\\
q_i
\end{pmatrix}
\end{equation}   
where $p_i, q_i$ are coordinates of real vector from the aforementioned two-dimensional subspace relative to the basis $\{{\rm Re}\,u_i, {\rm Im}\,u_i\}$. Thus, matrix $-\Omega {\mathscr H}$ can not be diagonalized over $\mathbb R$ but can be transformed to the following simple ('canonical`) form
\begin{equation}
\begin{pmatrix}
\boxed{\begin{matrix}
0&\omega_{10}\\
-\omega_{10}&0
\end{matrix}}&&&0\\
&\boxed{\begin{matrix}
0&\omega_{20}\\
-\omega_{20}&0
\end{matrix}}&&\\
&&\ddots&\\
0&&&\boxed{\begin{matrix}
0&\omega_{n0}\\
-\omega_{n0}&0
\end{matrix}}
\end{pmatrix}
\end{equation}

Note that spectrum of matrix $-\Omega {\mathscr H}$ is purely imaginary without dependence on character of critical point of the energy function. Identification of the index of the critical point under consideration can be performed by means of analysis of inequalities $|\lambda_i| +E_0>E_0$. Phase curves of Hamiltonian systems behave differently for critical points of different index. In particular, the solution of linearized Hamiltonian system reasonably approximates the solution of the initial non-linear system if  the Hessian of the Hamiltonian function is sign-definite (see, e.g., \cite{Arnold}). 

It is pertinent to mention that  by a certain abuse of notation we did not distinguish operator $H$ defined on the space $\mathbb C^{n+1}$ and its restriction on the subspace $\mathbb C^n$ complementary to $\mathbb Ce_0$.

Concluding this section we can state that for the projective spaces it is of no consequence what structure, Euclidean or symplectic, is used (if, of course, we are interested only in stability of energy critical points and excitation spectra but not in actual evolution). In the next sections it will be demonstrated that for submanifolds of projective spaces the situation is different: use of symplectic structure may give results essentially different from that obtained with the Euclidean structure.

\bigbreak \bigbreak {\Large \bf Hartree-Fock Manifolds} \bigbreak

We start with relevant assertions from the multilinear algebra. Their
proof may be found, e.g., in \cite{Sternberg, Greub, Kost}. Our presentation is close to that in \cite{Kost}. 
The notion of the wedge product is supposed to be known.

Symbol ${\cal F}_N^1$ will stand for 1-electron sector of the Fock space spanned by $n=|N|$ molecular spin orbitals (MSOs) $\{\psi_i\}$ with indices from the MSO index set $N=\{1,\ldots,n\}$. Vectors  from $ {\cal F}_N^p=\bigwedge^p{\cal F}_N^1$ are called $p-$vectors by mathematicians and $p-$electron states by physicists. $p$-vector $z$ is called decomposable if there exist  vectors $z_1,\ldots,z_p\in {\cal F}_N^1$ such that $z=z_1\wedge \ldots \wedge z_p$. In quantum chemistry decomposable $p-$vectors are called '$p-$electron Slater determinants`. Interpretation of quantum chemical notions in terms of modern multilinear algebra may be found in \cite{Cassam}.

\begin{prop} Vectors $z_1,z_2,\ldots,z_p$ from ${\cal
F}_N^1$ are linearly independent if and only if
\begin{equation}z_1\wedge \ldots \wedge z_p\ne 0.\end{equation}
\end{prop}

\begin{definition} For arbitrary $p-$vector $z$ its annihilator is 
\begin{equation} {\rm Ann}(z)=\{y\in {\cal F}_N^1|z\wedge y=0\}
\label{Ann}
\end{equation}
\end{definition}
It is clear that for any $p-$vector $z$ its annihilator is a subspace of the 
one-electron vector space ${\cal F}_N^1$. 
\begin{prop} Let $z$  be  $p-$vector with annihilator spanned by free vectors $z_1,z_2,\ldots,z_q$. Then 
there exists $(p-q)-$ vector $y$ such that
\begin{equation}
z=z_1\wedge z_2\wedge\ldots\wedge z_q\wedge y
\end{equation}
\end{prop}
In quantum chemistry annihilator of some $p-$electron state is called 'subspace of inactive MSOs associated with this state`. 
\begin{prop} Let $z_1,z_2,\ldots,z_p$ and
$y_1,y_2,\ldots,y_p$ be two free families of vectors from ${\cal
F}_N^1$. Then
\begin{equation}z_1\wedge \ldots \wedge z_p=\lambda y_1\wedge \ldots \wedge
y_p\ \ (\lambda \ne 0)\end{equation} if and only if $p-$planes (subspaces)
generated by vectors $z_1,z_2,\ldots,z_p$ and $y_1,y_2,\ldots,y_p$ are
identical.
\end{prop}
Since ${\rm Ann}(z)={\rm Ann}(\lambda z)$ for any $p-$vector $z$ and any $\lambda\ne 0$, it is possible to consider $\rm Ann$ as a mapping
\begin{equation}
{\rm Ann}:(\mathbb K \backslash \{0\})z_1\wedge z_2 \wedge \ldots \wedge z_p\to [z_1,z_2,\ldots,z_p]
\label{annmap}
\end{equation}
where  
\begin{equation}
[z_1,z_2,\ldots,z_p]=\sum\limits_{i=1}^p\mathbb Kz_i
\end{equation}
is a $\mathbb K -$linear hull of vectors $z_1,z_2,\ldots,z_p$.
 \begin{definition} The set of all
$p-$dimensional subspaces ($p-$planes) of one-electron Fock space  ${\cal F}_N^1$
is called its Grassmann manifold and is denoted by the
symbol ${\sf G}_p({\cal F}_N^1)$.  
\end{definition}
It is easy to see that the mapping defined by Eq.(\ref{annmap}) is actually a bijection and its inverse is an embedding of the Grassmann manifold ${\sf G}_p({\cal F}_N^1)$ into the 
the projective space $\mathbb P({\cal F}_N^p)$ of $p-$electron states.
\begin{definition}
The set ${\rm Ann}^{-1}({\sf G}_p({\cal F}_N^1))$ is called the Hartree-Fock (HF) manifold of $p-$electron states.
\end{definition}
HF manifold can be characterized implicitly as the set of solutions of a system of homogeneous polynomial equations, that is as a projective algebraic variety (see, e.g., \cite{Cox}). This characterization is based on the following simple statement (see, e.g., \cite{Kost}).
\begin{prop} For any non-zero $p-$vector the dimension of its annihilator is less or equal $p$. Non-zero $p-$vector is decomposable if and only if its annihilator is of dimension  $p$.    
\end{prop}
Recasting the vector equation
\begin{equation}
z\wedge y=0,\ z\in {\cal F}_N^p,\ y\in {\cal F}_N^1 
\end{equation}
in a coordinate form (with respect to some fixed one-electron basis), we arrive at a homogeneous linear system of $\binom{n}{p+1}$ scalar equations with respect to $n$ unknowns $y_1,y_2,\ldots,y_n$, and it is easy to see that $p-$vector $z$ is decomposable if and only if all minors of order $n-p+1$ of the matrix of this system are equal to zero. These conditions give us the required system of polynomial equations. 

Explicit characterization of HF manifolds in terms of local coordinates seems to be  much more useful for applications.

Let us fix some one-electron basis set $\{\psi_1,\psi_2,\ldots,\psi_n\}$ and consider $\mathbb K$-linear hull of vectors 
\begin{equation}
z_i=\sum\limits_{j=1}^n \psi_j Z_{ji}, \ i=1,2,\ldots,p
\end{equation}   
To this linear hull the mapping ${\rm Ann}^{-1}$ puts into correspondence the line generated by decomposable $p-$vector \begin{equation}
z=z_1\wedge\ldots\wedge z_p=\sum\limits_{1\le i_1<\ldots <i_p \le n}
\psi_{i_1}\wedge\ldots\wedge\psi_{i_p} Z_{i_1\ldots i_p}
\end{equation}
where $Z_{i_1\ldots i_p}$ is the determinant of $p\times p -$submatrix of $n\times p -$matrix $\rm Z$ with row indices $i_1,\ldots,i_p$. 

Coordinate charts on Grassmann manifold may be introduced as follows. Let us suppose that $Z_{i_1\ldots i_p}\ne 0$. Then we can write 
\begin{equation}(z_1\ldots ,z_p)=
(\psi_{i_1},\ldots,\psi_{i_p},\psi_{j_1},\ldots,\psi_{j_{n-p}})\left
(\begin{array}{c}
{\rm Z}_1\\
{\rm Z}_2 \end{array}\right ) \end{equation} where ${\rm Z}_1$ is
non-degenerate $p\times p-$submatrix of matrix ${\rm Z}$ with row indices $i_1,\ldots,i_p$ and ${\rm Z}_2$ is its $(n-p)\times p-$submatrix with complementary row indices $j_1,\ldots,j_{n-p}$. It is clear that vectors $(z_1\ldots ,z_p){\rm Z}_1^{-1}$
generate the same $p-$plane $[z_1,\ldots ,z_p]$. If another set 
\begin{equation}(y_1\ldots ,y_p)=
(\psi_{i_1},\ldots,\psi_{i_p},\psi_{j_1},\ldots,\psi_{j_{n-p}})\left
(\begin{array}{c}
{\rm Y}_1\\
{\rm Y}_2 \end{array}\right ) \end{equation}
of free vector generating this plane is chosen then necessarily ${\rm Y}_2{\rm Y}_1^{-1}={\rm Z}_2{\rm Z}_1^{-1}$. Thus, as a local parametrization  it is possible to take the mapping    
\begin{equation}
{\gamma}_{i_1i_2\ldots i_p}:{\rm Z}\to (\psi_{i_1},\ldots,\psi_{i_p},\psi_{j_1},\ldots,\psi_{j_{n-p}})\left
(\begin{array}{c}
{\rm I}_p\\
{\rm Z} \end{array}\right )\to [z_1\wedge z_2\wedge\ldots\wedge z_p] 
\label{grpar1}
\end{equation}
where ${\rm Z}\in {\mathbb K}^{p(n-p)}$ 
and  
\begin{equation}
z_k=\psi_{i_k}+\sum\limits_{l=1}^{n-p}\psi_{j_l}Z_{j_l k},\ (k=1,2,\ldots,p)
\label{zk}
\end{equation}
The domain of the corresponding chart is  $U_{i_1i_2 \ldots i_p}={\gamma}_{i_1i_2\ldots i_p}\left ({\mathbb K}^{p(n-p)}\right )$. It is easy to see that the family of charts 
\begin{equation}
 c_{i_1i_2\ldots i_p}=\left ( U_{i_1i_2 \ldots i_p},\varphi_{i_1i_2 \ldots i_p},{\mathbb K}^{p(n-p)} \right )
\end{equation}
where $\varphi_{i_1i_2 \ldots i_p}={\gamma}_{i_1i_2 \ldots i_p}^{-1}$, forms an atlas of Grassmann manifold and that this manifold is $\mathbb K -$analytic. Note as well that the chart with indices $i_1,\ldots,i_p$ is centered at the point ($p-$plane) $[\psi_{i_1},\ldots,\psi_{i_p}]$. It is pertinent to mention that the described atlas of the Grassmann manifold depends on the choice of the MSO basis set and by a properly selected non-degenerate transformation of MSOs any point of this manifold can be placed at the center of the 'standard` chart $c_{12\ldots p}=(U_{12\ldots p},\varphi_{12\ldots p},{\mathbb K}^{p(n-p)})$. For $p=1$ Eq.(\ref{grpar1}) is identical to Eq.(\ref{projpar1}). 
 
  In addition to the aforementioned 'canonical` realization of the Grassmann manifolds, there exist another realizations, of which we mention three most commonly used ones.  
   
(1) The set of all Hermitian idempotents (density operators) $\rho$ over ${\cal F}_N^1$ such that $Tr\ \rho=p$; 

(2) Quotient of the general linear group ${\rm GL}({\cal F}_N^1)$ modulo its certain closed subgroup.
 
(3) Quotient of the unitary group ${\rm U}({\cal F}_N^1)$ modulo its certain closed subgroup.

Two last realization require additional explanations. There is a natural transitive action of the general linear group ${\rm GL}({\cal F}_N^1)$  on  the set of all $p-$planes from  ${\cal F}_N^1$. If $\pi \in {\sf G}_p({\cal F}_N^1)$ then its isotropy group $G_{\pi}=\{g\in {\rm GL}({\cal F}_N^1)|g\pi=\pi\}$ is  a closed subgroup of ${\rm GL}({\cal F}_N^1)$. If $p$ first vectors of a chosen one-electron basis generate $p-$plane $\pi$ then matrix representation of transformation $g\in G_{\pi}$ is of the form
\begin{equation} 
\left(\begin{array}{cc}
\rm A&\rm B\\
0&\rm C \end{array}\right )
\label{Gomega}
\end{equation} 
The quotient space ${\rm GL}({\cal F}_N^1)/G_{\pi}$ can be endowed with $\mathbb K-$analytic structure consistent with the quotient topology \cite{Goto}. For any $p-$plane $\pi$ the set 
${\rm GL}({\cal F}_N^1)/G_{\pi}$ of left cosets of the general linear group ${\rm GL}({\cal F}_N^1)$ relative to $G_{\pi}$ is a homogeneous space isomorphic to the homogeneous space ${\sf G}_p({\cal F}_N^1)$. Indeed, let us put  
\begin{equation}
\theta_{\pi}:g{\rm G}_{\pi}\to g\pi
\end{equation}
We have 

(i) $\theta_{\pi}(g'g{\rm G}_{\pi})=g'g\pi=g'(g\pi)=g'\theta_{\pi}(g{\rm G}_{\pi})$ for any non-degenerate transformation $g'$ from ${\rm GL}({\cal F}_N^1)$;

(ii) $\theta_{\pi}(g{\rm G}_{\pi})=\theta_{\pi}(g'{\rm G}_{\pi})\Rightarrow g\pi=g'\pi\Rightarrow g{\rm G}_{\pi}=g'{\rm G}_{\pi}$;

(iii) For any $\pi '\in {\sf G}_p({\cal F}_N^1)$ there exists $g\in {\rm GL}({\cal F}_N^1)$ such that $\pi '=g\pi\Rightarrow \pi '=\theta_{\pi}(g{\rm G}_{\pi})$.

Property (i) means that the mapping $\theta_{\pi}$ is a morphism of homogeneous spaces. Properties (ii) and (iii) imply that this mapping is a bijection. 

Thus, for any fixed $p-$plane $\pi$ the Grassmann manifold ${\sf G}_p({\cal F}_N^1)$ can be identified with the space
\begin{equation} 
{\rm GL}({\cal F}_N^1)/{\rm G}_{\pi}=\bigcup\limits_{g\in {\rm GL}({\cal F}_N^1)\backslash {\rm G}_{\pi}}gG_{\pi}
\end{equation} 
of left cosets of the general linear group of one-electron Fock space relative to subgroup $G_{\pi}$. 

The general linear group ${\rm GL}({\cal F}_N^1)$ is a Lie group (that is a group and $\mathbb K-$analytic manifold) and its tangent space $\mathfrak g({\cal F}_N^1)={\sf T}_{I_n}({\rm GL}({\cal F}_N^1))$ at the identity $I_n$ is a Lie algebra of all one-electron linear transformations that is 
\begin{equation}
\mathfrak g({\cal F}_N^1)={\cal F}_N^1\otimes \left ({\cal F}_N^1\right )^*
\end{equation}
Subgroup $G_{\pi}$ is also a Lie group and its Lie algebra $\mathfrak g(\pi)$ is a subspace of the vector space $\mathfrak g({\cal F}_N^1)$. And again, if $p$ first vectors of a chosen one-electron basis generate $p-$plane $\pi$ then Lie algebra $\mathfrak g(\pi)$ can be identified with the algebra of matrices of the form of Eq.(\ref{Gomega}) with elements from the number field under consideration.  

Let us consider a decomposition
\begin{equation}
\mathfrak g({\cal F}_N^1)=\mathfrak g(\pi)\oplus \mathfrak p
\end{equation}
where subspace $\mathfrak p$ is constituted by matrices of the form
\begin{equation} 
\left(\begin{array}{cc}
0&0\\
\rm Z&0 \end{array}\right )
\end{equation} 
It can be proved (see, e.g., \cite{Goto}) that there exists a neighborhood of zero in the parameter space where the mapping
\begin{equation}
{\gamma}_{\pi}:{\rm Z}\to \exp\left ( {\rm Z} \right)G_{\pi}
\label{expg}
\end{equation}
with
\begin{equation}
{\rm Z}=\sum\limits_{\mu=1}^p\sum\limits_{\nu=p+1}^nZ_{\nu\mu}e_{\nu\mu}
\label{GLZ}
\end{equation}
and 
\begin{equation}
e_{\nu\mu}=|\psi_{\nu}\rangle\langle\psi_{\mu}|
\end{equation}
is a $\mathbb K-$analytic parametrization of a neighborhood of $G_{\pi}$ in ${\rm GL}({\cal F}_N^1)/{\rm G}_{\pi}$. The family of mappings $\{{\gamma}_{\pi}\}_{\pi}$ involves infinite number of members. Due to compactness of ${\rm G}_p({\cal F}_N^1)$  there exists a finite subfamily of this family that parametrizes this manifold. If $\pi$ is some $p-$plane spanned by MSOs  $\psi_1,\ldots,\psi_p$ then the mapping
\begin{equation}
\gamma_{12\ldots p}^{\rm GL}:{\rm Z}\to \left [\exp({\rm Z})\psi_1\right ]\wedge\ldots\wedge\left [\exp({\rm Z})\psi_p\right ]
\label{Gexp}
\end{equation}
may be used to parametrize representatives of HF lines in the projective space of
$p-$electron states belonging to a neighborhood of $p-$plane $\pi$.     

Both mappings (\ref{grpar1}) and  (\ref{Gexp}) are $\mathbb K-$analytic. But they have a certain drawback. Namely, even if the set of MSO corresponding to the  origin is orthonormal (with respect to the standard Euclidean or Hermitian scalar product on one-electron Fock space), the parametrized $p-$frames corresponding to non-zero values of parameters are not. This may be inconvenient both for evaluation of matrix elements and in the course of solution of optimization problem. It is easy to modify the definition  of the mapping (\ref{grpar1}) to   eliminate the aforementioned drawback: \begin{equation}
\gamma_{12\ldots p}^{\rm GS}:{\rm Z}\to (\psi_{1},\ldots,\psi_{n}){\rm g(Z)} \to \left [{\rm g(Z)}\psi_{1}\right ]\wedge\ldots \wedge \left [{\rm g(Z)}\psi_{p}\right ] 
\label{GS}
\end{equation}
where
\begin{equation}
{\rm g(Z)}=\left(\begin{array}{c}
{\rm I}_p\\
{\rm Z} \end{array}\right ){\rm W(Z)}
\label{gZ} 
\end{equation}
and
${\rm W(Z)}$ is $p\times p$
upper triangle matrix performing Gram-Schmidt orthogonalization of
vectors representing  $p-$plane from ${U_{12 \ldots p}}$, and where the initial one-electron basis is
supposed to be orthonormal ( as has been already mentioned, without loss of generality it is possible to consider $p-$planes from  ${U_{12 \ldots p}}$). For $p=1$ the mapping (\ref{GS}) coincides with the mapping (\ref{reppar}). For the case of the complex parameter space of dimension $p(n-p)$ its  realification  leads to the real  parameter space of dimension $2p(n-p)$ with  basis $\{e_{\nu\mu},{\rm i}e_{\nu\mu}\}$. This basis is orthonormal with respect to the Euclidean scalar product $\rm (\rm Z,Z')={\rm Re}\ [{\rm Tr}\ Z^{\dagger}Z']$ and symplectic with respect to skew-symmetric scalar product $\rm [Z,Z']={\rm Im}\ [{\rm Tr}\ Z^{\dagger}Z']$.

To modify properly the definition of the parametrization (\ref{Gexp}), it is necessary to consider the unitary subgroup  ${\rm U}({\cal F}_N^1)$ of the general linear group. Unitary  transformations possess the following important properties: (1) unitary group acts on the set of all $p-$planes transitively, and (2) if $p-$plane $\pi$ is invariant with respect to $u\in {\rm U}({\cal F}_N^1)$ then orthogonal complement $\pi^{\perp}$ is also invariant with respect to $u$. As a result, the isotropy group of arbitrary $p-$plane $\pi$ is a direct product $G_{\pi}={\rm U}(\pi)\times {\rm U}(\pi^{\perp})$. Lie algebra $\mathfrak u({\cal F}_N^1)$ of the unitary group consists of skew-Hermitian matrices with, in general,  complex elements but it is a vector space over the field $\mathbb R$ of real numbers (after multiplication by, say, the imaginary unit  skew-Hermitian matrix becomes Hermitian). The parameter space for the exponential parametrization  should be taken as the space of all matrices of the form    
\begin{equation}
{\rm Z}=\sum\limits_{\nu=p+1}^n\sum\limits_{\mu=1}^p\left [Z_{\nu\mu}e_{\nu\mu}-{\bar Z}_{\nu\mu}e_{\mu\nu}\right ]
\label{UZ}
\end{equation}
and parametrization mapping  $\gamma_{12\ldots p}^{\rm U}$ is given by  Eq.(\ref{Gexp}) but with parameter matrix (\ref{UZ}) instead of the matrix (\ref{GLZ}). 
 
Orthogonal with respect to the trace inner product basis in this real parameter space can be chosen as
\begin{subequations}
\begin{eqnarray}
a_{\nu\mu}=&e_{\nu\mu}-e_{\mu\nu}\\
s_{\nu\mu}=&{\rm i}\left (e_{\nu\mu}+e_{\mu\nu}\right )
\end{eqnarray}
\label{bas}
\end{subequations}
In this basis the parameter matrix (\ref{UZ}) takes the form
\begin{equation}
{\rm Z}=\sum\limits_{\nu=p+1}^n\sum\limits_{\mu=1}^p\left [X_{\nu\mu}a_{\nu\mu}+Y_{\nu\mu}s_{\nu\mu}\right ]
\end{equation}
where $\rm X,Y$ are real and imaginary components of $(n-p)\times p$ parameter matrix $\rm Z$.

Symplectic form on real even-dimensional parameter space with basis $\{a_{\nu\mu},s_{\nu\mu}\}$  is defined with the aid of coordinate functionals $dX_{\nu\mu}:{\rm Z}\to X_{\nu\mu}$ and $dY_{\nu\mu}:{\rm Z}\to Y_{\nu\mu}$ as
\begin{equation} 
\omega=\sum\limits_{\nu=p+1}^n\sum\limits_{\mu=1}^p dX_{\nu\mu}\wedge dY_{\nu\mu}
\label{sform}
\end{equation}

If $\gamma_{12\ldots p}(\rm Z) $ is a local parametrization of  representatives of HF states (built with orthonormal MSOs) then  $2p(n-p)$ vectors 
\begin{subequations}
\begin{eqnarray}
\frac{\partial {\gamma}_{12\ldots p}}{\partial X_{\nu \mu}}(\rm Z)=\sum\limits_{j=1}^p z_1\wedge \ldots \wedge
\frac{\partial z_j}{\partial X_{\nu \mu}}\wedge \ldots
\wedge z_p\\
\frac{\partial {\gamma}_{12\ldots p}}{\partial Y_{\nu \mu}}(\rm Z)=\sum\limits_{j=1}^p z_1\wedge \ldots \wedge
\frac{\partial z_j}{\partial Y_{\nu \mu}}\wedge \ldots
\wedge z_p
\end{eqnarray}
\end{subequations}
form a basis of the tangent space to the HF manifold at the point $\gamma_{12\ldots p}(\rm Z)$. 

We start with the energy functional parametrized by $\gamma^{\rm U}_{12\ldots p}(\rm Z)$: 
\begin{equation}
E({\rm Z})=\frac{1}{2}\langle \gamma^{\rm U}_{12\ldots p}({\rm Z})|H|\gamma^{\rm U}_{12\ldots p}({\rm Z}) \rangle
\label{EHF}
\end{equation}
where $H$ is a Hermitian operator acting on the $p-$electron sector ${\cal F}_N^p$ of the Fock space. We suppose that this functional reaches its minimum at the origin $\rm X=Y=0$ of the parameter space and the corresponding single-determinant wave function is $\Phi_0=\psi_1\wedge\ldots\wedge\psi_p$.  

When using exponential parametrization $\gamma^{\rm U}_{12\ldots p}(Z)$,  we confine ourselves to the quadratic approximation of energy functional and, consequently, to the linearized version of the Hamiltonian equations. 

Local expansion of $\gamma^{\rm U}_{12\ldots p}({\rm Z})$ in a neighborhood of the origin is
\begin{equation}
\gamma^{\rm U}_{12\ldots p}({\rm Z})=\left [1-\frac{1}{2}{\rm Tr}\ ({\rm Z}^{\dagger}{\rm Z})\right ]\Phi_0+\sum\limits_{\lambda,i}\Phi_i^{\lambda}Z_{\lambda i}+\sum\limits_{\lambda,\lambda '}\sum\limits_{i<i'}
\Phi_{ii'}^{\lambda \lambda'}Z_{\lambda i}Z_{\lambda' i'}+\cdots
\end{equation} 
From this expansion it readily follows that basis of the tangent space to the HF manifold at the origin is constituted by  $2p(n-p)$ 'single excited determinants':
\begin{equation}
\frac{\partial \gamma^{\rm U}_{12\ldots p}}{\partial X_{\nu  \mu}}(0)=\Phi_{\mu}^{\nu},\quad 
\frac{\partial \gamma^{\rm U}_{12\ldots p}}{\partial Y_{\nu\mu}}(0)={\rm i}\Phi_{\mu}^{\nu}
\label{sebas}
\end{equation}
Energy differential at the origin is 
\begin{eqnarray}
d_{(0,0)}E=\frac{1}{2}\sum\limits_{\nu=p+1}^n\sum\limits_{\mu=1}^p\left [\langle \Phi_{\mu}^{\nu}|H|\Phi_0\rangle +\langle \Phi_0|H|\Phi_{\mu}^{\nu}\rangle \right ]dX_{\nu\mu}\nonumber \\
+\frac{1}{2}\sum\limits_{\nu=p+1}^n\sum\limits_{\mu=1}^p\left [\langle {\rm i}\Phi_{\mu}^{\nu}|H|\Phi_0\rangle +\langle \Phi_0|H|{\rm i}\Phi_{\mu}^{\nu}\rangle \right ]dY_{\nu\mu}
\end{eqnarray}
and it is easy to see that stationary conditions $d_{(0,0)}E(a_{\nu\mu})=d_{(0,0)}E(s_{\nu\mu})=0$ are equivalent to the well-known in quantum chemistry Brillouin conditions $\langle \Phi_{\mu}^{\nu}|H|\Phi_0\rangle =0$. Note that we have not specified yet the concrete Hermitian operator involved in Eq.(\ref{EHF}) and, consequently, even the classic form of the  Brillouin theorem is of a rather general nature.

At this stage it is convenient to introduce the following matrices:
\begin{subequations}
\begin{eqnarray}
{\rm H}^{\rm CIS}_{\nu\mu,\nu'\mu'}&=&\langle \Phi_{\mu}^{\nu}|H|\Phi_{\mu'}^{\nu'}\rangle\\
\Delta_{\nu\mu,\nu'\mu'}&=&(1-\delta_{\nu\nu'})(1-\delta_{\mu\mu'})\langle \Phi_0|H|\Phi_{\mu\mu'}^{\nu\nu'}\rangle
\end{eqnarray}
\end{subequations}
Note that ${\rm H}^{\rm CIS}$ is Hermitian whereas $\Delta$ is symmetric.
 
Differential of the quadratic part of  energy function (\ref{EHF}) at a point $\rm (X,Y)$ of the parameter space is
\begin{equation}
d_{{\rm (X,Y)}}E^{(2)}=\begin{pmatrix} d{\rm X}\ d{\rm Y}\end{pmatrix}\mathscr H\begin{pmatrix} {\rm X}\\{\rm Y}\end{pmatrix}
\end{equation}
where 
\begin{equation}
\mathscr H=\begin{pmatrix} {\rm Re}\left (\rm H^{\rm CIS}+ \Delta\right)-E_0{\rm I}_{p(n-p)}&-{\rm Im}\left (\rm H^{\rm CIS}+ \Delta\right )\\
{\rm Im}\left (\rm H^{\rm CIS}- \Delta\right )&{\rm Re}\left (\rm H^{\rm CIS}- \Delta\right)-E_0{\rm I}_{p(n-p)}
\end{pmatrix}
\label{hessmatrix}
\end{equation}
is the matrix of the energy function Hessian calculated at the origin of the parameter space with respect to the basis (\ref{bas}). Here $E_0=\langle \Phi_0|H|\Phi_0\rangle=2E(0)$.

The next step is to use the symplectic form (\ref{sform}) to get the linearized Hamiltonian equations analogous to 
Eqs.(\ref{matHequ}) for the case of the projective manifolds. We have
\begin{equation}
\left (\begin{array}{c}
\overset{.}{\rm X}\\
\overset{.}{\rm Y}
\end{array}\right )=
-\Omega \mathscr H \left (\begin{array}{c}
{\rm X}\\
{\rm Y}
\end{array}\right )
\label{matHequ1}
\end{equation}
where
\begin{eqnarray}
\Omega=\left (\begin{array}{cc}
0&{\rm I}_{p(n-p)}\\
-{\rm I}_{p(n-p)}&0
\end{array}\right )
\end{eqnarray}
is the matrix of the symplectic form (\ref{sform}).  

Complexification of the parameter space  gives $2p(n-p)$-dimensional complex space where the matrix Hamiltonian equation (\ref{matHequ1}) takes the form
\begin{eqnarray}
\left (\begin{array}{c}
\overset{.}{\rm Z}\\
\overset{.}{\bar {\rm Z}}
\end{array}\right )=
{\rm i}\left (\begin{array}{cc}
\ {\rm H}^{\rm CIS}-E_0{\rm I}_{p(n-p)}&\bar\Delta\\
-\Delta&\ -(\bar{{\rm H}}^{\rm CIS}-E_0{\rm I}_{p(n-p)})
\end{array}\right )
\left (\begin{array}{c}
{\rm Z}\\
\bar {\rm Z}
\end{array}\right )
\label{Shrod3}
\end{eqnarray}

At this stage we have three matrices: real symmetric matrix of the second order derivatives $\mathscr H$, real Hamiltonian matrix $-\Omega\mathscr H$, and  complex Schr$\rm \ddot {o}$dinger matrix $-{\rm C}^{-1}\Omega\mathscr H\rm C$ where
\begin{equation}
{\rm C}=\frac{1}{2}\begin{pmatrix}
{\rm I}_{p(n-p)}&{\rm I}_{p(n-p)}\\
-{\rm i}{\rm I}_{p(n-p)}&{\rm i}{\rm I}_{p(n-p)}
\end{pmatrix}
\end{equation} 
First two matrices are relative to the basis (\ref{bas}) of real parameter space and the third matrix is relative to $\rm C-$transformed basis $\{1\otimes a_{\nu\mu},1\otimes s_{\nu\mu}\}$ of the complexified parameter space. Using isomorphism  $d_{(0,0)}\gamma^U_{12\ldots p}$, we can identify the parameter space with the tangent space to the HF manifold at the point $\gamma^U_{12\ldots p}(0,0)$. With such an identification matrices $\mathscr H$ and $-\Omega\mathscr H$ can be considered relative to the basis of single excited determinants (\ref{sebas}) whereas Schr$\rm \ddot {o}$dinger matrix becomes relative to the orthogonal basis 
\begin{subequations}
\begin{eqnarray}
\Upsilon_{\mu}^{\nu}=\frac{1}{2}\left [(1\otimes \Phi_{\mu}^{\nu})-{\rm i}(1\otimes {\rm i}\Phi_{\mu}^{\nu})\right ]\\
\bar{\Upsilon}_{\mu}^{\nu}=\frac{1}{2}\left [(1\otimes \Phi_{\mu}^{\nu})+{\rm i}(1\otimes {\rm i}\Phi_{\mu}^{\nu})\right ]
\end{eqnarray}  
\label{csebas}
\end{subequations}
of the complexified tangent space.

Matrix $\mathscr H$ of the second order derivatives is orthogonally diagonalized over $\mathbb R$ and has real spectrum consisting of $2p(n-p)$, in general case different, eigenvalues. Hamiltonian matrix $\Omega\mathscr H$ obviously satisfies the equality $(\Omega\mathscr H)^t\Omega+\Omega(\Omega\mathscr H)=0$ which means that it is an infinitesimal-symplectic matrix (element of Lie algebra of the symplectic group $\rm Sp(\mathbb R^{2n})$). It is easy to show that spectrum of such (non-degenerate) matrix satisfies the following restrictions: (i) if $\lambda$ is its eigenvalue then necessarily $-\lambda$ is also its eigenvalue; (ii) each complex eigenvalue  appears in pair with its conjugate.  And of course, Hamiltonian and Schr$\rm \ddot {o}$dinger matrices  have identical spectra. These matrices, however, are neither Hermitian nor skew-Hermitian. They are not, in general, orthogonally diagonalized. In contrast to the case of the projective space, spectrum of these matrices is not necessarily purely imaginary. 

Till now $p-$electron Hermitian operator $H$ was not specified and all equations obtained are valid for arbitrary choice of this operator. For electronic Hamiltonian all matrix elements involved, say, in Eq.(\ref{Shrod3}), may be easily calculated with the aid of the standard technique to give
\begin{subequations}
\begin{eqnarray}
{\rm H}^{\rm CIS}_{\nu\mu,\nu'\mu'}&=& \left [E_0+\varepsilon_{\nu}-\varepsilon_{\mu}\right ]\delta_{\mu\mu'}\delta_{\nu\nu'}+\langle \nu\mu'\|\mu\nu'\rangle\\
\Delta_{\nu\mu,\nu'\mu'} &=&\langle \mu\mu'\|\nu\nu'\rangle
\end{eqnarray}
\end{subequations}
where $\langle ij\|kl\rangle=\langle ij|kl\rangle -\langle ij|lk\rangle$.
Substitution of these expressions in Eq.(\ref{Shrod3}) immediately gives the standard TDHF equations (see, e.g., \cite{Thouless}). 

In analogy with classical mechanics, matrix $\Delta$ in Eqs.(\ref{matHequ1})-(\ref{Shrod3}) may be interpreted as a constraining matrix  and its appearance is of the same nature as appearance of constraining force in Newton equations. The case $\Delta =0$ corresponds to CIS method which reduces to diagonalization of operator $H$ projection  on the subspace of $p-$electron states spanned by vectors $\{\Phi_{\mu}^{\nu}\}$.

As we have already seen, choice of Euclidean or symplectic metric on the projective space leads, in essence, to equivalent theories. For HF manifold, however, it is not the case. And in our opinion there is no {\it a priroi} criterion, what metric is preferable. Strictly speaking, it is necessary to compare  behavior of exact excitations energies, TD excitation energies, and Hessian eigenvalues as functions of (complex) parameters $\Delta _{\nu\mu,\nu'\mu'}$ to make reliable conclusion. In Appendix A  analysis of behavior of excitation energies is performed for a simple model case.

Now let us turn to Gram-Schmidt parametrization (\ref{GS}). Using Eqs. (B.15)-(B.16) from Appendix B, it is easy to get the following general expressions for partial derivatives:
\begin{subequations}
\begin{eqnarray}
\frac{\partial }{\partial X_{\nu\mu}}\gamma_{12\ldots p}^{\rm GS}({\rm Z})=\sum\limits_{j=1}^p [{\rm W(Z)}]_{\mu j}\Phi_{j}^{\nu}({\rm Z})
-{\rm Re}\left [{\rm g(Z)W}^{\dagger}({\rm Z})\right ]_{\nu\mu}\Phi({\rm Z})\\
\frac{\partial }{\partial Y_{\nu\mu}}\gamma_{12\ldots p}^{\rm GS}({\rm Z})=\sum\limits_{j=1}^p [{\rm W(Z)}]_{\mu j}{\rm i}\Phi_{j}^{\nu}({\rm Z})
+{\rm i}\,{\rm Im}\left [{\rm g(Z)W}^{\dagger}({\rm Z})\right ]_{\nu\mu}{\rm i}\Phi({\rm Z})
\end{eqnarray} 
\end{subequations}
where 
\begin{eqnarray}
\Phi ({\rm Z})=z_1\wedge z_2\wedge \ldots \wedge z_p=\gamma_{12\ldots p}^{\rm GS}({\rm Z})\qquad\ \\
\Phi_{j}^{\nu}({\rm Z})=z_1\wedge\ldots\wedge z_{j-1}\wedge \psi_{\nu}\wedge z_{j+1}\ldots\wedge z_p
\end{eqnarray}
Energy derivatives are easily calculated at arbitrary point $\rm Z=X+i Y$. In realified parameter space $\mathbb R^{2p(n-p)}$ symplectic scalar product may be introduced as the imaginary part of the Hermitian trace product. After transformation of energy differential to symplectic gradient it is possible to write down {\it the exact} Hamiltonian equations:
\begin{subequations}
\begin{eqnarray}
\overset{.}X_{\nu\mu}=2{\rm Im}\left [\sum\limits_{j=1}^pW_{\mu j}({\rm Z})\langle \Phi({\rm Z})|H|\Phi_{j}^{\nu}({\rm Z})\rangle +\left ({\rm g(Z)W}^{\dagger}({\rm Z})\right )_{\nu\mu}E({\rm Z})\right ]\\
\overset{.}Y_{\nu\mu}=2{\rm Re}\left [\sum\limits_{j=1}^pW_{\mu j}({\rm Z})\langle \Phi({\rm Z})|H|\Phi_{j}^{\nu}({\rm Z})\rangle -\left ({\rm g(Z)W}^{\dagger}({\rm Z})\right )_{\nu\mu}E({\rm Z})\right ]
\end{eqnarray}
\end{subequations}
where $E({\rm Z})=\frac{1}{2}\langle \Phi ({\rm Z})|H|\Phi ({\rm Z})\rangle $.
 
The corresponding Schr$\rm \ddot{o}$dinger-type equations are
\begin{subequations}
\begin{eqnarray}
\overset{.}Z_{\nu\mu}= 2{\rm i}\left [\sum\limits_{j=1}^p {\overline W}_{\mu j}({\rm Z})\overline {\langle \Phi({\rm Z})|H|\Phi_{j}^{\nu}({\rm Z})\rangle} - \left ({\rm g(Z)W}^{\dagger}({\rm Z})\right )_{\nu\mu}E({\rm Z})\right ]\\
\overset{.}{\bar Z}_{\nu\mu}=-2{\rm i}\left [\sum\limits_{j=1}^p W_{\mu j}({\rm Z})\langle \Phi({\rm Z})|H|\Phi_{j}^{\nu}({\rm Z})\rangle - \overline{\left ({\rm g(Z)W}^{\dagger}({\rm Z})\right )}_{\nu\mu}E({\rm Z})\right ]
\end{eqnarray} 
\end{subequations}
Remind once again that these equations are exact (not linearized) ones. Their linearization will not lead to new equations, because it does not depend on the choice of local parametrization (see, e.g., \cite{Arnold}).
 
The most common in quantum theory approach uses Grassmann manifold without its prior embedding into  $p$-electron projective space. This approach is applicable to general energy functionals that can be defined in terms of 1-density idempotent operators. Since exponential parametrization of Grassmann manifolds can be found (in explicit or implicit form) in almost all publications concerning TD theories (see, e.g.,\cite{Thouless, Rowe-1}), we confine ourselves to analysis of Gram-Schmidt parametrization. Instead of the mapping (\ref{GS}) we have 
\begin{equation}
\pi_{12\ldots p}^{\rm GS}:{\rm Z}\to \sum\limits_{k=1}^p |z_k\rangle\langle z_k| 
\label{GS1}
\end{equation}      
where $z_k={\rm g(Z)}\psi_{k}$ and $\rm g(Z)$ is given by Eq.(\ref{gZ}). To simplify notations, the explicit indication on dependence of matrices $\rm g$ and $\rm W$ on parameters $\rm X,Y$ will be omitted. For the same reason we suppress subscript  $12\ldots p$ (chart multiindex) and superscript GS of the parametrization mapping (\ref{GS1}). 

Expressions for partial derivatives of the mapping $\pi({\rm X,Y})$ are easily derived from Eqs.(B.6)-(B.9) of Appendix B:
\begin{subequations}
\begin{eqnarray}
\frac{\partial}{\partial X_{\nu\mu}}\pi({\rm X,Y})=\sum\limits_{k=1}^p\left[W_{\mu k}|\psi_{\nu}\rangle\langle z_k|+
\overline{W}_{\mu k}|z_k\rangle\langle\psi_{\nu}|\right ]-\nonumber\\
\sum\limits_{k,l=1}^p\left [W_{\mu l}{\bar g}_{\nu k}+\overline {W}_{\mu k}g_{\nu l}\right ]|z_k\rangle\langle z_l|\qquad\\
\frac{\partial}{\partial Y_{\nu\mu}}\pi({\rm X,Y})={\rm i}\sum\limits_{k=1}^p\left[W_{\mu k}|\psi_{\nu}\rangle\langle z_k|-
\overline{W}_{\mu k}|z_k\rangle\langle\psi_{\nu}|\right ]-\nonumber\\
{\rm i}\sum\limits_{k,l=1}^p\left [W_{\mu l}{\bar g}_{\nu k}-\overline {W}_{\mu k}g_{\nu l}\right ]|z_k\rangle\langle z_l|\qquad
\end{eqnarray}
\label{denpar}
\end{subequations}
In particular, real tangent space to the Grassmann manifold at the origin is spanned by the vectors
\begin{subequations}
\begin{eqnarray}
\frac{\partial}{\partial X_{\nu\mu}}\pi(0,0)=& e_{\nu\mu}+e_{\mu\nu}\\
\frac{\partial}{\partial Y_{\nu\mu}}\pi(0,0)=&{\rm i}(e_{\nu\mu}-e_{\mu\nu})
\end{eqnarray}
\label{1stder}
\end{subequations}
where $e_{\nu\mu}=|\psi_{\nu}\rangle\langle\psi_{\mu}|$, $\mu=1,\ldots ,p$, and $\nu=p+1,\ldots,n$. 

The second order derivatives of the parametrization mapping at the origin are also easily calculated to give
\begin{subequations}
\begin{eqnarray}
\frac{\partial ^2}{\partial X_{\nu\mu}\partial X_{\nu'\mu'}}\pi(0,0)=
\delta_{\mu\mu'}(e_{\nu\nu'}+e_{\nu'\nu})-\delta_{\nu\nu'}(e_{\mu\mu'}+e_{\mu'\mu})\quad\\
\frac{\partial ^2}{\partial X_{\nu\mu}\partial Y_{\nu'\mu'}}\pi(0,0)=-{\rm i}\delta_{\mu\mu'}(e_{\nu\nu'}-e_{\nu'\nu})-{\rm i}\delta_{\nu\nu'}(e_{\mu\mu'}-e_{\mu'\mu})\\
\frac{\partial ^2}{\partial Y_{\nu\mu}\partial Y_{\nu'\mu'}}\pi(0,0)=\frac{\partial ^2}{\partial X_{\nu\mu}\partial X_{\nu'\mu'}}\pi(0,0)\qquad\qquad
\end{eqnarray}
\label{2ndder}
\end{subequations}

Let us suppose that orthonormal MSO basis is fixed and each 1-electron operator is identified with its matrix. In particular, Grassmann manifold ${\sf G}_p({\cal F}_N^1)$ can be identified with $n\times n-$matrices $\rho$ satisfying the following restrictions:
\begin{equation}
\rho^{\dagger}=\rho,\, {\rm Tr}\,\rho =p,\, \rho^2=\rho
\end{equation}

A function $E(\rho)$,   
smooth with respect to real variables $\alpha_{ij}={\rm Re}\,\rho_{ij},\beta_{ij}={\rm Im}\,\rho_{ij}$ and, in general, complex-valued, will be referred to as 'the energy function`, or just 'the energy'.  
Within the chart under consideration (with multiindex $12\ldots p$) local representative of the energy function restriction to the Grassmann manifold is $\mathscr E({\rm X,Y})=E\circ\pi({\rm X,Y})$.
It is reasonable to suppose that for physically relevant energy functions the imaginary component of $\mathscr E({\rm X,Y})$ vanishes (note that, even if this condition is fulfilled, the imaginary part of $E(\rho)$ should not necessarily be equal to zero for arbitrary complex matrix $\rho$).  To avoid cumbersome expressions, we suppose that the energy is a holomorphic function of complex variables $\rho_{ij}$. In this case realification of the energy domain is not required.     
   
Using Eqs.(\ref{denpar}), it is easy to calculate partial derivatives of $\mathscr E({\rm X,Y})$ at arbitrary point $({\rm X,Y})$ of the parameter space:
\begin{subequations}
\begin{eqnarray}
\frac{\partial \mathscr E}{\partial X_{\nu\mu}}({\rm X,Y})=\left [\left (\rm {I}_n-{\rm gg}^{\dagger}\right ){\frac{d{\bar E}}{d\rho}}^{\dagger}\rm gW^{\dagger}\right ]_{\nu\mu}+\left [{\rm Wg}^{\dagger}{\frac{d{\bar E}}{d\rho}}^{\dagger}\left (\rm {I}_n-{\rm gg}^{\dagger}\right )\right ]_{\mu\nu}\\
\frac{\partial \mathscr E}{\partial Y_{\nu\mu}}({\rm X,Y})={\rm i}\left [\left (\rm {I}_n-{\rm gg}^{\dagger}\right ){\frac{d{\bar E}}{d\rho}}^{\dagger}\rm gW^{\dagger}\right ]_{\nu\mu}-{\rm i}\left [{\rm Wg}^{\dagger}{\frac{d{\bar E}}{d\rho}}^{\dagger}\left (\rm {I}_n-{\rm gg}^{\dagger}\right )\right ]_{\mu\nu}
\end{eqnarray}
\end{subequations}
where $\frac{dE}{d\rho}$ is, in general complex, matrix of partial derivatives $\frac{\partial E}{\partial \rho_{ij}}$ calculated at the point $\pi({\rm X,Y})$. 

The Schr$\rm \ddot{o}$dinger-type evolution equations on the complexified parameter space are
\begin{subequations}
\begin{eqnarray}
\overset{.}Z_{\nu\mu} &= \ \ 2\rm i\, \left [{\rm Wg}^{\dagger}{\frac{d{\bar E}}{d\rho}}^{\dagger}\left (\rm {I}_n-{\rm gg}^{\dagger}\right )\right ]_{\mu\nu} \\
\overset{.}{\bar Z}_{\nu\mu} &= -2\rm i\, \left [\left (\rm {I}_n-{\rm gg}^{\dagger}\right ){\frac{d{\bar E}}{d\rho}}^{\dagger}\rm gW^{\dagger}\right ]_{\nu\mu} 
\end{eqnarray}
\label{schrgras}
\end{subequations}

To linearize  these equations in a neighborhood of the origin it is necessary to calculate the first and the second derivatives of $\mathscr E({\rm X,Y})$ at the point $(0,0)$. It can be easily done with the aid of Eqs.(\ref{1stder})-(\ref{2ndder}). The linearized  Schr$\rm \ddot{o}$dinger-type equations are
\begin{subequations}
\begin{eqnarray}
\overset{.}Z_{\nu\mu} ={\rm i}\sum\limits_{\nu'\mu'}\left [\frac{\partial^2 E}{\partial \rho_{\mu\nu}\partial \rho_{\nu'\mu'}}+\delta_{\mu\mu'}\frac{\partial E}{\partial \rho_{\nu'\nu}}-\delta_{\nu\nu'}\frac{\partial E}{\partial \rho_{\mu\mu'}}\right ]Z_{\nu'\mu'}\nonumber\\
+{\rm i}\sum\limits_{\nu'\mu'}\frac{\partial^2 E}{\partial \rho_{\mu\nu}\partial \rho_{\mu'\nu'}}{\bar Z}_{\nu'\mu'}\qquad\qquad\qquad\\
\overset{.}{\bar Z}_{\nu\mu} =-{\rm i}\sum\limits_{\nu'\mu'}\left [\frac{\partial^2 E}{\partial \rho_{\nu\mu}\partial \rho_{\mu'\nu'}}+\delta_{\mu\mu'}\frac{\partial E}{\partial \rho_{\nu\nu'}}-\delta_{\nu\nu'}\frac{\partial E}{\partial \rho_{\mu'\mu}}\right ]{\bar Z}_{\nu'\mu'}\nonumber\\
-{\rm i}\sum\limits_{\nu'\mu'}\frac{\partial^2 E}{\partial \rho_{\nu\mu}\partial \rho_{\nu'\mu'}}Z_{\nu'\mu'}\qquad\qquad\qquad
\end{eqnarray}
\label{schrgrasl}
\end{subequations}
Derivatives on the right-hand side of these equations are taken at the point $\rho_0=\left (\begin{smallmatrix}{\rm I}_p&0\\0&0\end{smallmatrix}\right )$. 

Classic HF theory supplies us with an example of simple energy function that can be defined in two ways: 
\begin{equation}
E^{(1)}_{\rm HF}(\rho)=\frac{1}{2}{\rm Tr}\, \rho\left [{\rm h}+ {\rm F}(\rho)\right ]
\end{equation}   
or 
\begin{equation}
E^{(2)}_{\rm HF}(\rho)=\frac{1}{2}{\rm Tr}\, \rho^{\dagger}\left [{\rm h}+ {\rm F}(\rho)\right ]
\end{equation}   
where 
\begin{equation}
F_{ij}(\rho)=h_{ij}+\sum\limits_{k,l}\rho_{lk}\langle ik\|jl\rangle
\end{equation}
These functions coincide on the subspace of Hermitian matrices $\rho$ but different as functions on the space of all complex matrices. In particular, $E^{(1)}_{\rm HF}(\rho)$ is holomorphic as a function of complex variables $\rho_{ij}$ whereas $E^{(2)}_{\rm HF}(\rho)$ is not. 

It is easy to show that Fock matrix ${\rm F}(\rho)$ is Hermitian for any Hermitian $\rho$ and that  the energy $E^{(1)}_{\rm HF}$ (and, consequently, $E^{(2)}_{\rm HF}$) restriction to the Grassmann manifold is a real-valued function.  

We confine ourselves to the holomorphic case. Simple calculation give
\begin{subequations}
\begin{eqnarray}
\frac{\partial E^{(1)}_{\rm HF}}{\partial \rho_{ij}}(\rho)=F_{ji}(\rho)\\
\frac{\partial^2 E^{(1)}_{\rm HF} }{\partial \rho_{ij}\partial \rho_{kl}}(\rho)=\langle jl\|ik \rangle
\end{eqnarray}
\end{subequations}
where, within the chosen chart,  $\rho({\rm X,Y})\equiv\pi({\rm X,Y})$. 

The Schr$\rm \ddot{o}$dinger-type evolution equations on the complexified parameter space, corresponding to the energy function $E^{(1)}_{\rm HF}(\rho)$, are is readily obtained from Eq.(\ref{schrgras})
\begin{subequations}
\begin{eqnarray}
\overset{.}Z_{\nu\mu} &= \ \ 2\rm i\, \left [Wg^{\dagger}F(\rho)\left ( I_{\it  n}-gg^{\dagger}\right )\right ]_{\mu\nu}\ \\
\overset{.}{\bar Z}_{\nu\mu} &= -2\rm i\,\left [Wg^{\dagger}F(\rho)\left ( I_{\it  n}-gg^{\dagger}\right )\right ]_{\nu\mu}
\end{eqnarray}
\end{subequations}

Using formulas (\ref{schrgrasl}a)-(\ref{schrgrasl}b), it is easy to derive the linearized version of these equations:
\begin{subequations}
\begin{eqnarray}
\overset{.}Z_{\nu\mu}={\rm i}\sum\limits_{\nu'\mu'}\left [\langle \nu\mu'\|\mu\nu' \rangle+\delta_{\nu\nu'}\delta_{\mu\mu'}(\varepsilon_{\nu}-\varepsilon_{\mu})\right ]Z_{\nu'\mu'}\nonumber\\
+{\rm i}\sum\limits_{\nu'\mu'}\langle \nu\nu'\|\mu\mu' \rangle {\bar Z}_{\nu'\mu'}\qquad\qquad\\
\overset{.}{\bar Z}_{\nu\mu}=-{\rm i}\sum\limits_{\nu'\mu'}\left [\langle \mu\nu'\|\nu\mu' \rangle+\delta_{\nu\nu'}\delta_{\mu\mu'}(\varepsilon_{\nu}-\varepsilon_{\mu})\right ]{\bar Z}_{\nu'\mu'}\nonumber\\
-{\rm i}\sum\limits_{\nu'\mu'}\langle \mu\mu'\|\nu\nu' \rangle Z_{\nu'\mu'}\qquad\qquad
\end{eqnarray}
\end{subequations}
where it is assumed that the origin of the parameter space is placed at a stationary point of $\mathscr E^{(1)}_{\rm HF}({\rm X,Y})$ and the canonical HF MSOs corresponding to this point are selected.

The energy $E^{(2)}_{\rm HF}$ (non-holomorphic case) should be treated either as a function of real variables $\alpha_{ij}$ and $\beta_{ij}$ or  complex variables $\rho_{ij}$ and ${\bar \rho}_{ij}$.  
\bigbreak

{\Large \bf Conclusion  }

\bigbreak

Physical TD theories, put properly in the framework of modern differential geometry, may become  a general and powerful tool for investigation of many electron systems. As soon as general scheme of derivation of evolution equations in the case of, say, relatively simple complex projective spaces is    elaborated, the same scheme with minor technical modifications can be applied for any projective algebraic manifold. Of course, existence of additional mathematical structures on the manifold under consideration may result in   plenty of different equivalent realization of the same geometrical object and each realization may carry its own unique feature. Selection of relevant realization depends, of course, on the concrete  task. In present paper we considered complex Grassmann manifold and its realizations as (1) a submanifold of  many electron projective space which we called HF manifold, and (2) the set of idempotent 1-density operators.  In our opinion, both realizations supplement each other. Study of another realizations exploiting the fact that Grassmann manifolds are also homogeneous spaces of certain Lie groups may be found in \cite{Rowe-1}-\cite{Rowe-3}.       

Comparison of linearized Schr$\rm \ddot{o}$dinger evolution equations on many electron projective space and on its HF submanifold readily reveals the appearance of a constraining matrix $\Delta$ in the HF case. This matrix includes explicitly double excitations from the HF state and characterizes in a certain sense the curvature of the HF manifold. Perturbation analysis of eigenvalues (TD transition energies) of the matrix of linearized Schr$\rm \ddot{o}$dinger evolution equation on HF manifold leads to the conclusion that these eigenvalues are concave functions of matrix elements of the constraining matrix whereas the exact transition energies are usually convex functions of the same parameters. This means that, if a given CIS transition energy is greater than the exact one then, at least in a neighborhood of $\Delta =0$, TD transition energy should be more close to the exact value than the CIS one. This feature of TD transition energies is not so clearly seen if realization of Grassmann manifold as the set of 1-density idempotents is used.

\bigbreak        
{\Large \bf Acknowledgments}
\bigbreak
 The author  gratefully acknowledges the Russian Foundation for
Basic Research (Grant 06-03-33060) for financial support of the
present work.

\bigbreak \bigbreak
{\Large \bf Appendix A. Simple Example  of  }

{\Large\bf Excitation Energies Behavior} 
\bigbreak \bigbreak

Let us consider simple case corresponding to the projection of $p-$electron operator $H$ on the subspace spanned by determinants 
\begin{equation}
\Phi_0, \Phi_{\mu}^{\nu}, \Phi_{\mu\mu'}^{\nu\nu'}\, \nonumber
\end{equation}
where $\Phi_0=\psi_1\wedge\psi_2\wedge\ldots\wedge\psi_p$. General form of relevant operator $H$ matrix in this case is
$$
{\rm H}(\Delta)=
\left (\begin{array}{ccc}
E_0&0&\Delta \\
0&{\rm H}^{\rm CIS}&{\rm A}\\
\Delta^{\dagger}&{\rm A}^{\dagger}&{\rm B}
\end{array}\right )
\eqno(A.1)$$
where $\Delta_{\nu\mu,\nu'\mu'}$ is a matrix of free complex parameters, $\rm A$ and $\rm B$ are arbitrary fixed complex and Hermitian matrices, respectively.
  
Writing matrix $\rm H(\Delta)$ in the form
$$
{\rm H}_{\varepsilon}(\Delta)={\rm H}(0)+\varepsilon{\rm W}(\Delta)
\eqno(A.2)
$$
where 
$$
{\rm W}(\Delta)=\sum\limits_{\nu\nu'}\sum\limits_{\mu\mu'}\left [{\Delta}_{\nu\mu,\nu'\mu'}|\Phi_0\rangle\langle\Phi_{\mu\mu'}^{\nu\nu'}|+\bar{\Delta}_{\nu\mu,\nu'\mu'}|\Phi_{\mu\mu'}^{\nu\nu'}\rangle\langle \Phi_0|\right ]
\eqno(A.3)
$$
and treating ${\rm W}(\Delta)$ as perturbation, we come to the following second order expression for the exact transition energies:
$$
\omega_{i0}(\Delta)=\omega_{i0}(0)+\sum\limits_{j\ne 0}\frac{|\sum\limits_{\nu\nu'}\sum\limits_{\mu\mu'}\bar{\Delta}_{\nu\mu,\nu'\mu'}\langle\Psi_j|\Phi_{\mu\mu'}^{\nu\nu'}\rangle|^2}{\omega_{j0}(0)}(1+\delta_{ji})
\eqno(A.4)
$$   
where $\Psi_j$ are the eigenvectors of matrix ${\rm H}(0) (\Psi_0=\Phi_0)$. It is clear that $\omega_{i0}(\Delta)$ are convex quadratic functions.   

Matrix of the Schr$\rm \ddot{o}$dinger-type equation (\ref{Shrod3}) (the imaginary unit prefactor is omitted)  with respect to the basis (\ref{csebas}) of the complexified tangent space may be written in the form (A.2) with
$$
{\rm H}(0)=
\left (\begin{array}{cc}
{\rm H^{CIS}}-E_0{{\rm I}_{p(n-p)}}&0\\
0&-({\rm {\bar H}^{CIS}}-E_0{{\rm I}_{p(n-p)}}
\end{array}\right )
\eqno(A.5)$$
and (non-Hermitian) perturbation
$$
W(\Delta)=2Q({\bar \Delta})-2Q({\bar \Delta})^{\dagger}
\eqno(A.6)
$$
where 
$$
Q({\bar \Delta})=\sum\limits_{\nu\nu'}\sum\limits_{\mu\mu'}{\bar \Delta}_{\nu\mu,\nu'\mu'}|\Upsilon_{\mu}^{\nu}\rangle\langle\bar{\Upsilon}_{\mu'}^{\nu'}|\eqno(A.7)
$$
Prefactor 2 in Eq.(A.6) appears due to somewhat non-standard normalization of basis functions (\ref{csebas}): $\langle\Upsilon_{\mu}^{\nu}|\Upsilon_{\mu}^{\nu}\rangle=\langle\bar{\Upsilon}_{\mu}^{\nu}|\bar{\Upsilon}_{\mu}^{\nu}\rangle=\frac{1}{2}$.
 
Zero-order matrix $\rm H(0)$ has real spectrum $\pm \omega^{\rm TD}_{i0}(0)$ symmetric with respect to zero. If $\Psi_i$ is eigenvector of block (1,1) of $\rm H(0)$ corresponding to the eigenvalue $\omega^{\rm TD}_{i0}(0)$ (CIS transition energy) then ${\bar \Psi}_i$ is the eigenvector of block (2,2) of this matrix with the eigenvalue $-\omega^{\rm TD}_{i0}(0)$. We confine ourselves to the perturbation analysis of non-negative part of matrix $\rm H(\Delta)$ spectrum. The second order expression for transition energies in this case is 
$$
\omega^{\rm TD}_{i0}(\kappa)=\omega^{\rm TD}_{i0}(0)-4\sum\limits_{j\ne i}\frac{|\langle \Psi_i|Q{\bar \Delta}|{\bar \Psi}_{j}\rangle |^2}{\omega^{\rm TD}_{i0}+\omega^{\rm TD}_{j0}}
\eqno(A.8)
$$
From (A.8) it readily follows that, at least in some neighborhood of $\Delta =0$,  TD transition energies are concave functions of parameters  $\Delta_{\nu\mu,\nu'\mu'}$. As a result, if for a fixed index $i$ graphics of functions $\omega_{i0}(\Delta)$ and $\omega^{\rm TD}_{i0}(\Delta)$ have crossing points, then at this points TD excitation energy takes on the exact value. If $\omega_{i0}^{\rm TD}(0)>\omega_{i0}(0)$ then there exists a certain neighborhood  of the point  $\Delta =0$ where TD transition energy $\omega_{i0}^{\rm TD}(\Delta)$ more close to the exact values than the CIS energy. Probably in this sense one should interpret frequently occurring in literature statement that TDHF approach is superior in many aspects than the CIS one (see, e.g. \cite{Rowe-1}).      

\begin{figure}[ht]
 \begin{center}
 		\includegraphics[totalheight=2.27in]{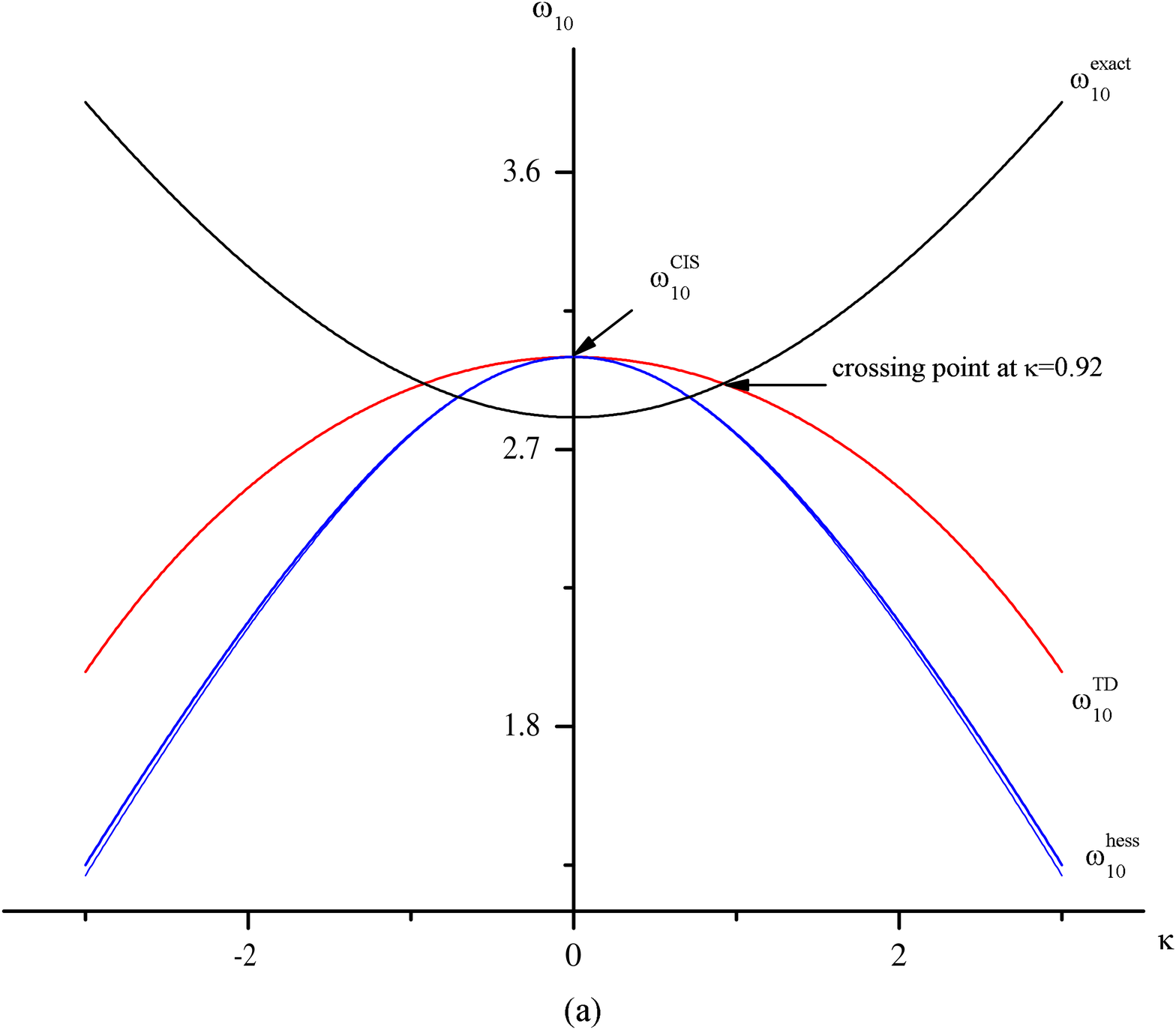}
 		\includegraphics[totalheight=2.27in]{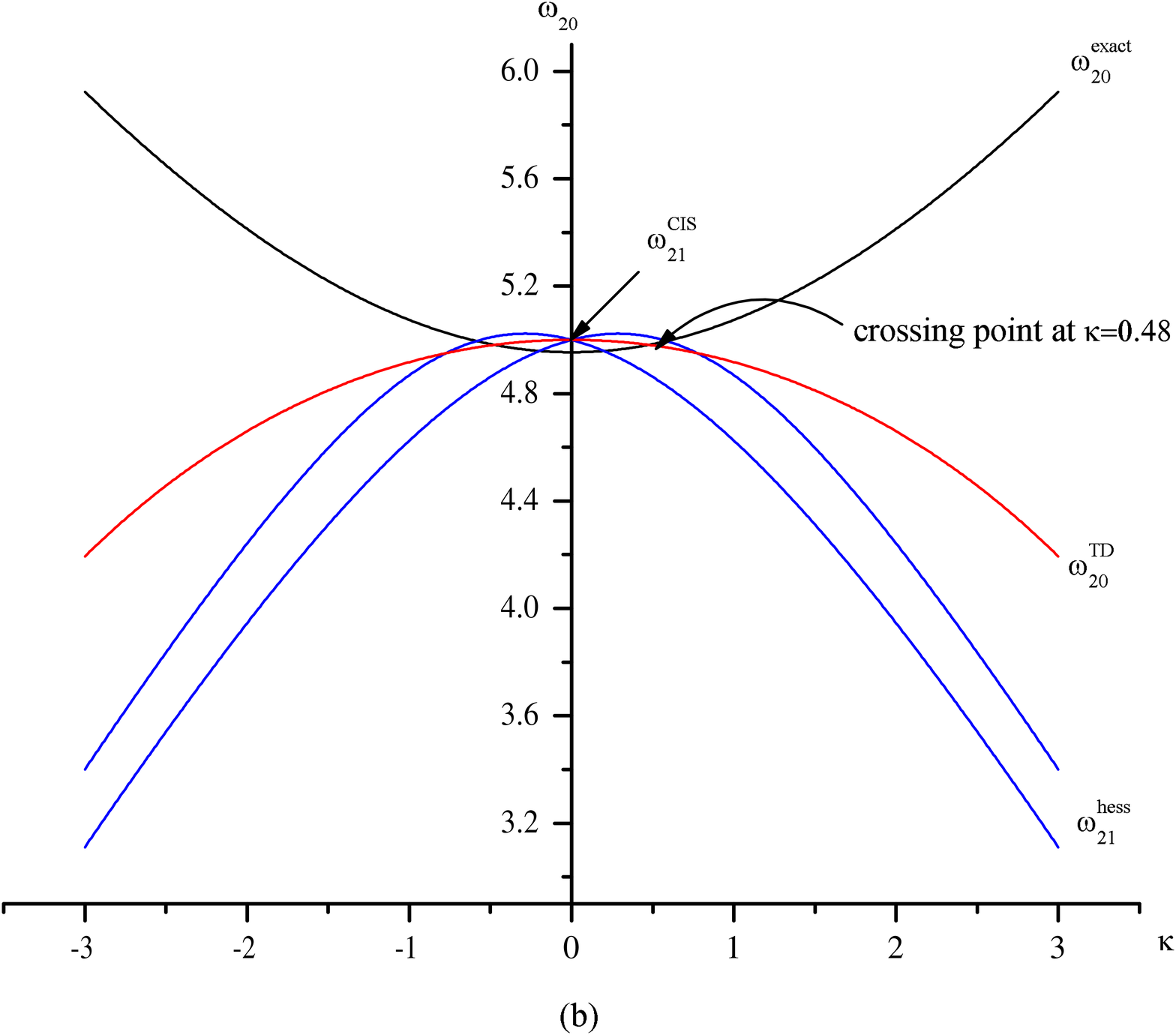}
 	\end{center}
 \begin{center}
 		\includegraphics[totalheight=2.27in]{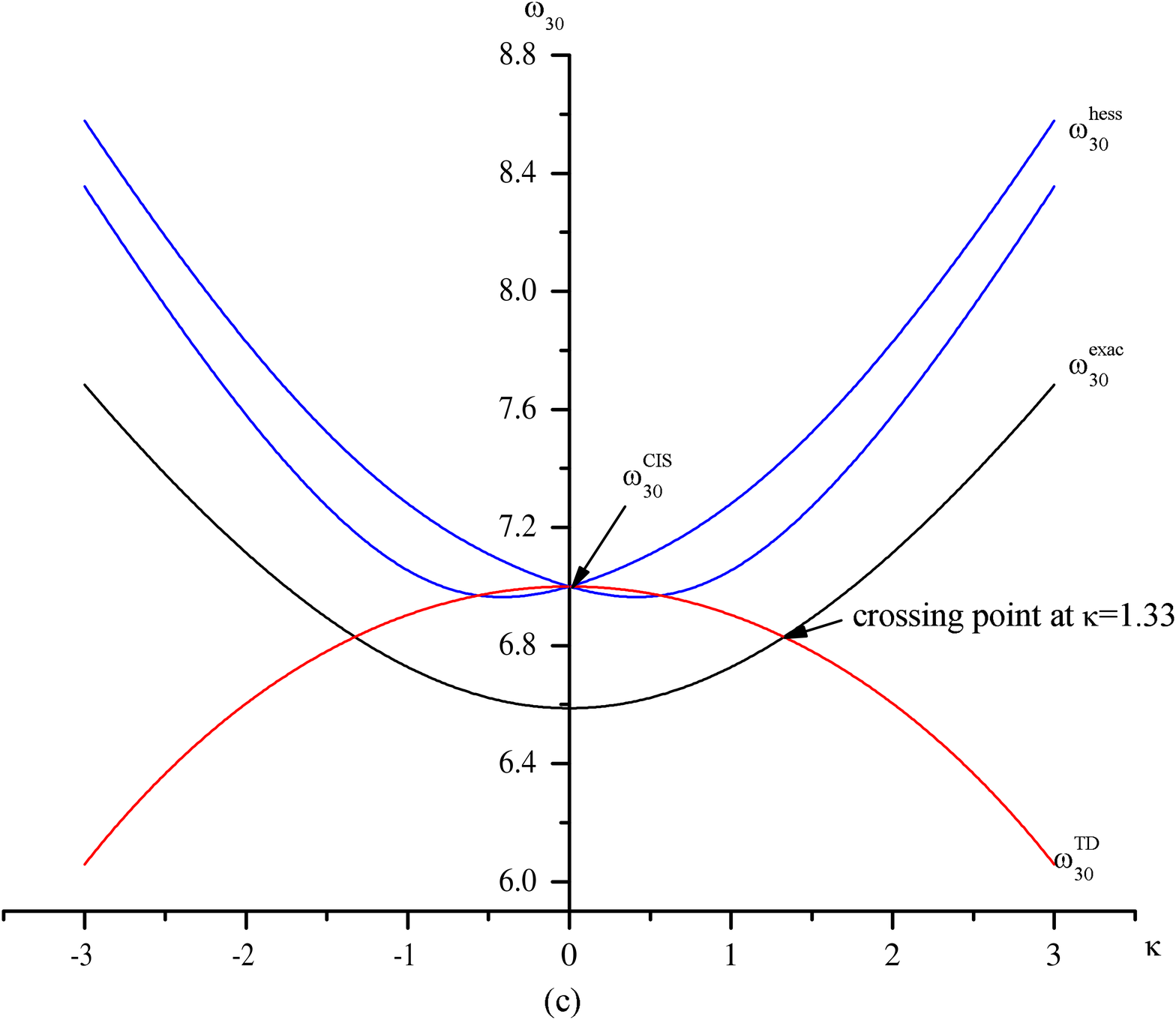}
 		\includegraphics[totalheight=2.27in]{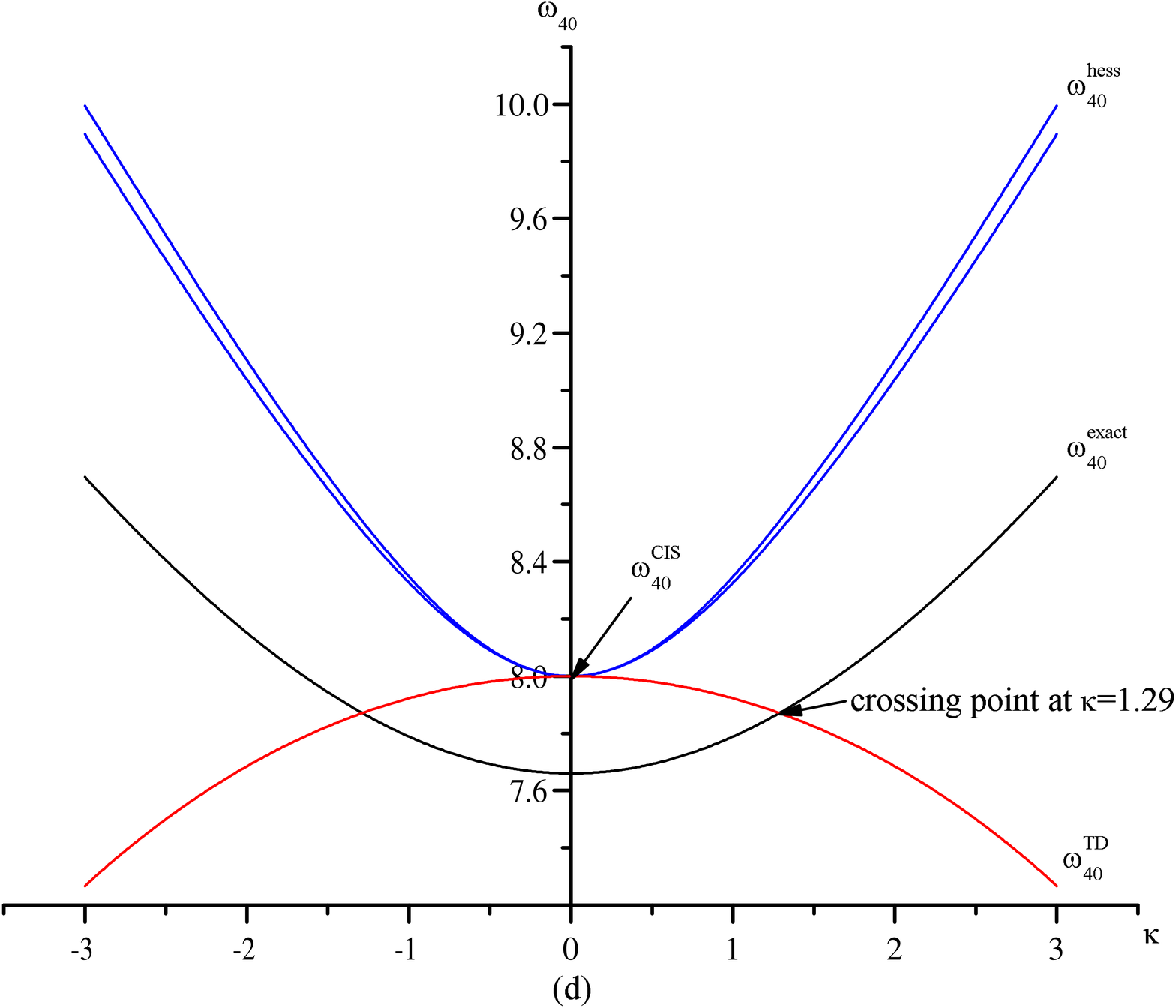}
 	\end{center}
 	\caption{Transition energies as functions of parameter $\kappa$}
 	\label{fig1}
\end{figure}

To visualize the exact dependence of transition energies on single parameter $\kappa$, let us take the concrete complex Hermitian $6\times 6$ matrix 
relative to the basis $\{\Phi_0, \Phi_1^3, \Phi_1^4, \Phi_2^3, \Phi_2^4, \Phi_{12}^{34}\}$
$$
{\rm H}(\kappa)=
\left (
\begin{smallmatrix}
15&0&0&0&0&\kappa\\
0&-11.8433&-0.31349 - 0.47024{\rm i}&0& 0.04167+ 0.1528{\rm i}&0.2+{\rm i}\\
0& -0.3135+ 0.47024{\rm i}&-9.9623&0& -0.54167 - 0.1806{\rm i}&0.3\\
0& 0&0&-7&0&0.1-{\rm i}\\
0&0.04167 - 0.1528{\rm i}&-0.5417+ 0.1806{\rm i}&0& -8.1944&0.2-{\rm i}\\
{\bar\kappa}&0.2-{\rm i}&0.3&0.1+{\rm i}&0.2+{\rm i}&-6
\end{smallmatrix}
\right )\eqno(A.9)
$$
and use graphical tools of the Mathematica package \cite{Wolfram}. In Fig.\ref{fig1} dependence of the exact transition energies, TD transition energies and Hessian eigenvalues on real parameter $\kappa$ is displayed. 

\begin{figure}[ht]
 \begin{center}
 		\includegraphics[totalheight=3.3in]{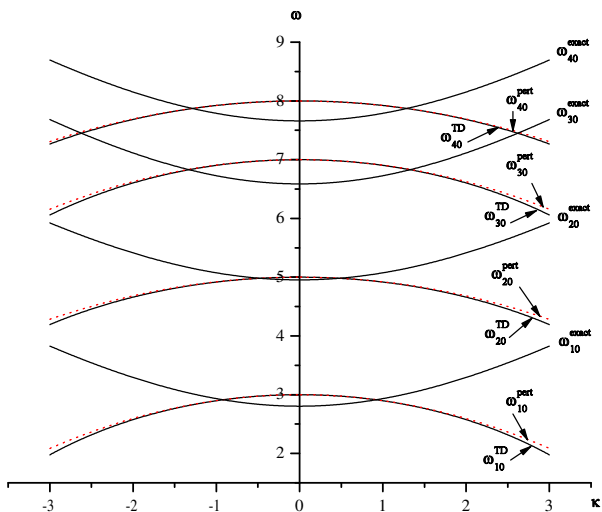}
 	\end{center}
 	\caption{Exact transition energies $\omega_{i0}^{\rm exact}$, TD transition energies $\omega_{i0}^{\rm TD}$, and second order TD transition energies $\omega_{i0}^{\rm pert}$  as functions of parameter $\kappa$}
 	\label{fig2}
\end{figure}

Exact transition energies are convex and TD transition energies are concave functions of $\kappa$ in  certain neighborhoods of the origin  for all four excited states. The aforementioned intersection points are essentially different for different excited states. For example, $\omega^{\rm TD}_{10}(\kappa)\xrightarrow[\kappa\to 0.92]{} \omega^{\rm exact}_{10}$ and $\omega^{\rm TD}_{30}(\kappa)\xrightarrow[\kappa\to 1.33]{} \omega^{\rm exact}_{30}$. 
In Fig.\ref{fig2} the second order approximations of TD transition energies are shown together with the exact and TD transition energies. 
It is seen that at least in the concrete case under consideration the second order perturbation correction (A.8) gives excellent approximation of TD transition energies for all states. 

In some cases use of perturbation theory to evaluate excitation energies may be computationally more efficient than the direct diagonalization of either non-symmetric Hamiltonian real matrix or non-Hermitian Schr$\rm \ddot{o}$dinger-type complex matrix. However, to make reliable conclusion about applicability of perturbation approach, heavy numerical testing is required.       
\bigbreak \bigbreak

\newpage
{\Large \bf Appendix B. Analytic Expressions for }

{\Large \bf Derivatives of Gram-Schmidt} 

{\Large \bf Parametrization Mapping  } 
\bigbreak \bigbreak

Matrix
$$
{\rm g(Z)}=\left (\begin{array}{c} {\rm I}_p\\
{\rm Z} \end{array}\right ){\rm W(Z)}
\eqno{(B.1)}$$ involved in Eq.(\ref{GS}) satisfies the orthogonality condition
$$
\left [{\rm g(Z)}\right ]^{\dagger}{\rm g(Z)}={\rm I}_p
\eqno{(B.2)}$$
To simplify the notations, we agree to omit, when possible, 
explicit  indication on dependence of matrices $\rm W$ and $\rm g$ on $\rm Z=X+iY$.
Differentiation of Eq.(B.2) gives
$$ \frac {\partial {\rm W}^{\dagger}}{\partial X_{\nu\mu}}
\left [{\rm W}^{\dagger}\right ]^{-1}+{\rm W}^{-1}\frac {\partial {\rm W}}{\partial
X_{\nu\mu}}=-{\rm W}^{\dagger}{\rm J}^{\dagger}_{\nu\mu}{\rm g}-{\rm g}^{\dagger}{\rm J}_{\nu\mu}{\rm W}\eqno{(B.3a)}$$
$$\frac {\partial {\rm W}^{\dagger}}{\partial Y_{\nu\mu}}
\left [{\rm W}^{\dagger}\right ]^{-1}+{\rm W}^{-1}\frac {\partial {\rm W}}{\partial
Y_{\nu\mu}}={\rm i}{\rm W}^{\dagger}{\rm J}^{\dagger}_{\nu\mu}{\rm g}-{\rm i}{\rm g}^{\dagger}{\rm J}_{\nu\mu}{\rm W}\quad
\eqno{(B.3b)}$$ 
where basis matrices ${\rm J}_{\nu\mu}$ and ${\rm i}{\rm J}_{\nu\mu}$ are defined by 
$$\frac {\partial}{\partial X_{\nu\mu}}\left (\begin{array}{c} {\rm I}_p\\
{\rm Z} \end{array}\right )={\rm J}_{\nu\mu}\eqno(B.4a)$$
$$\frac {\partial}{\partial Y_{\nu\mu}}\left (\begin{array}{c} {\rm I}_p\\
{\rm Z} \end{array}\right )={\rm i}{\rm J}_{\nu\mu}\eqno(B.4b)$$

Following Garton \cite{Garton}, let us introduce the matrices
$$
{\rm P}_{\nu\mu}={\rm W}^{-1}\frac {\partial {\rm W}}{\partial X_{\nu\mu}}\eqno(B.5a)$$
$${\rm Q}_{\nu\mu}={\rm W}^{-1}\frac {\partial {\rm W}}{\partial Y_{\nu\mu}}
\eqno{(B.5b)}$$
where $\nu=p+1,\ldots,n$ and $\mu =1,\ldots,p$. Since $\rm W$ is upper triangle, matrices ${\rm P}_{\nu\mu}$ and ${\rm Q}_{\nu\mu}$ are also upper triangle  and their matrix elements are
$$
\left [{\rm P}_{\nu\mu}\right ]_{ij}=\begin{cases}
\ \ \ 0 &\text{if\ } i>j \cr 
-\frac{1}{2}\left [{\rm W}^{\dagger}{\rm J}^{\dagger}_{\nu\mu}{\rm g}\right ]_{ii}-\frac{1}{2}\left [{\rm g}^{\dagger}{\rm J}_{\nu\mu}{\rm W}\right ]_{ii} &\text{if\ } i=j \cr
 -\left[{\rm W}^{\dagger}{\rm J}^{\dagger}_{\nu\mu}{\rm g}\right ]_{ij}-\left[{\rm g}^{\dagger}{\rm J}_{\nu\mu}{\rm W}\right
 ]_{ij} &\text{if\ } i<j\cr
 \end{cases}\eqno(B.6a)$$
$$\left [{\rm Q}_{\nu\mu}\right ]_{ij}=\begin{cases}
\ \ \ 0 &\text{if\ } i>j \cr 
\frac{\rm i}{2}\left [{\rm W}^{\dagger}{\rm J}^{\dagger}_{\nu\mu}{\rm g}\right ]_{ii}-\frac{\rm i}{2}\left [{\rm g}^{\dagger}{\rm J}_{\nu\mu}{\rm W}\right ]_{ii} &\text{if\ } i=j \cr
 {\rm i}\left[{\rm W}^{\dagger}{\rm J}^{\dagger}_{\nu\mu}{\rm g}\right ]_{ij}-{\rm i}\left[{\rm g}^{\dagger}{\rm J}_{\nu\mu}{\rm W}\right
 ]_{ij} &\text{if\ } i<j\cr
 \end{cases}\quad
 \eqno{(B.6b)}$$
 At the origin ${\rm P}_{\nu\mu}(0)={\rm Q}_{\nu\mu}(0)=0$.
 
 From Eqs.(B.5)-(B.6) it follows that
 $$
 \frac {\partial {\rm W}}{\partial X_{\nu\mu}}={\rm WP}_{\nu\mu}\eqno(B.7a)$$
$$\frac {\partial {\rm W}}{\partial Y_{\nu\mu}}={\rm WQ}_{\nu\mu}
 \eqno{(B.7b)}$$
 and
$$
 \frac {\partial {\rm g}}{\partial
 X_{\nu\mu}}={\rm J}_{\nu\mu}{\rm W}+{\rm gP}_{\nu\mu}\eqno(B.8a)$$
$$ \frac {\partial {\rm g}}{\partial
 Y_{\nu\mu}}={\rm i}{\rm J}_{\nu\mu}{\rm W}+{\rm gQ}_{\nu\mu}
 \eqno{(B.8b)}$$
 
 Now it is easy to calculate the derivatives of vectors $z_j={\rm g(Z)}\psi_j$
 with respect to parameters $X_{\nu\mu}$ and $Y_{\nu\mu}$:
 $$
 \frac {\partial z_j}{\partial
 X_{\nu\mu}}={\psi}_{\nu}W_{\mu j}+\sum\limits_{k=1}^{j} z_k\left
 [{\rm P}_{\nu\mu}\right ]_{kj}\eqno{(B.9a)}$$
 $$\frac{\partial z_j}{\partial
 Y_{\nu\mu}}={\rm i}{\psi}_{\nu}W_{\mu j}+\sum\limits_{k=1}^{j} z_k\left
 [{\rm Q}_{\nu\mu}\right ]_{kj}
 \eqno{(B.9b)}$$
At the origin we have
$$
 \frac {\partial {\rm W}}{\partial X_{\nu\mu}}(0)= \frac {\partial {\rm W}}{\partial Y_{\nu\mu}}(0)=0\eqno(B.10a)
$$
$$\frac {\partial {\rm g}}{\partial X_{\nu\mu}}(0)={\rm J}_{\nu\mu},\quad \frac {\partial {\rm g}}{\partial
 Y_{\nu\mu}}(0)={\rm i}{\rm J}_{\nu\mu}\eqno(B.10b)
 $$
 $$\frac {\partial z_j}{\partial X_{\nu\mu}}(0)={\psi}_{\nu}\delta_{\mu j},\quad
 \frac {\partial z_j}{\partial Y_{\nu\mu}}(0)={\rm i}{\psi}_{\nu}\delta_{\mu j}\eqno(B.10c)
 $$

 \bigbreak

\bigbreak

\end{document}